\documentclass[english,jfm]{jfm}
\usepackage[T1]{fontenc}
\usepackage[latin9]{inputenc}
\usepackage{verbatim}
\usepackage{amstext}
\usepackage{graphicx}
\usepackage{amssymb}
\usepackage[authoryear]{natbib}

\usepackage{babel}

\begin{document}

\title{The catalytic role of beta effect in barotropization processes}


\author[A. Venaille, G. K. Vallis and S. M. Griffies]%
{A.\ns V\ls E\ls N\ls A\ls I\ls L\ls L\ls E$^1$%
  \thanks{Present address: Laboratoire de Physique ENS-Lyon., 69007 Lyon, France, antoine.venaille@ens-lyon.org},\ns
G.\ns K.\ns V\ls A\ls L\ls L\ls I\ls S$^1$\break
\and S.\ns M.\ns G\ls R\ls I\ls F\ls F\ls I\ls E\ls S$^1$}

\affiliation{NOAA, GFDL, AOS Program, Princeton University, NJ O8540, USA.}

\date{\today}
\maketitle


\begin{abstract}	
The vertical structure of freely evolving, continuously stratified, quasi-geo\-strophic flow is investigated. We predict the final state organization, and in particular its vertical structure, using statistical mechanics and these predictions are tested against numerical simulations. The key role played by conservation laws in each layer, including the fine-grained enstrophy, is discussed. In general, the conservation laws, and in particular that enstrophy is conserved layer-wise, prevent complete barotropization, i.e., the tendency to reach the gravest vertical mode. The peculiar role of the $\beta$-effect, i.e. of the existence of planetary vorticity gradients, is discussed. In particular, it is shown that increasing $\beta$ increases the tendency toward barotropization through turbulent stirring. The effectiveness of barotropization may be partly parameterized using the Rhines scale $2\pi E_{0}^{1/4}/\beta^{1/2}$. As this parameter decreases (beta increases) then barotropization can progress further, because the beta term provides enstrophy to each layer.  However, if the beta effect is too large then the statistical mechanical predictions fail and wave dynamics prevent complete barotropization. 
\end{abstract}

\section{Introduction}

The vertical structure of eddying flow in the oceanic mesoscale is a fundamental
problem in geophysical fluid dynamics, one that has has been reinvigorated by the
need to interpret altimetric observation of surface velocity fields
\citep{ScottWang05,LapeyreJPO09}. Energy cascades and self-organization processes
are fairly well understood and characterized in two-dimensional turbulence, but
how these results generalize to rotating, continuously stratified flows is still
an open question. We ask in this paper what is the final organization of a freely
evolving inviscid quasi-geostrophic flow with continuous stratification. In
particular, we examine what are the precise conditions for the oft-cited
barotropization process to occur. Barotropization refers to the tendency of a
quasi-geostrophic flow to reach the gravest vertical mode \citep{Charney71,
Rhines77}. Because the gravest vertical mode is the barotropic one, i.e. a depth
independent flow, ``barotropization'' means a tendency toward the formation of
depth independent flows. We study in particular the key role played by the beta
effect (the existence of planetary vorticity gradients) in such barotropization
processes. It has been previously noticed that the presence of these large scale
planetary vorticity gradients favors barotropic process, see e.g. \citet{SmithVallis01}, but this phenomenon remains to our knowledge unexplored and unexplained.

Continuously stratified quasi-geostrophic flows take place in three dimensions,
but their dynamics are quasi-two-dimensional because the non-divergent advecting
velocity field has only horizontal components. The dynamics admits similar
conservation laws as the two-dimensional Euler equations, including among others
the total energy and the enstrophy of each layer. These conservation laws imply
an inverse energy cascade toward large scales, and weak energy dissipation in
presence of weak viscosity
\citep{Kraichnan_Motgommery_1980_Reports_Progress_Physics}. This situation
contrasts with three dimensional turbulence where the energy cascades forward
toward small scales, giving rise to finite energy dissipation no matter how weak
the viscosity.

Just as for two-dimensional Euler flows, inviscid, freely evolving continuously stratified quasi-geostrophic flows are disordered and an unstable initial condition rapidly evolves into a strongly disordered fine-grained flow with filaments being stretched, folded, and consequently cascading toward smaller and smaller scales. However, the inverse energy cascade leads to the formation of robust, large scale coherent structures filling the domain in which the flow takes place, contrasting strongly with the analogous three dimensional flow with its energy transfer to small scales.

The aim of this paper is to provide a prediction for the vertical structure of the large scale flow organization resulting from inviscid, freely evolving continuously stratified quasi-geostrophic dynamics using a statistical mechanics approach. The goal of the statistical mechanics approach is to drastically reduce the complexity of the problem of determining the large scale flow organization to the study of a few key parameters given by the dynamical invariants, as for instance the total energy and the fine-grained enstrophy in each layer. Such a theory has been independently proposed by \cite{Robert:1990_CRAS,SommeriaRobert:1991_JFM_meca_Stat} and \cite{Miller:1990_PRL_Meca_Stat}, and henceforth will be referred to as MRS theory. It is an equilibrium theory, developed in the case without forcing and dissipation. Our contribution will be to apply this theory to the explicit computation of a class of equilibrium states of the stratified quasi-geostrophic equations.

{A crucial ingredient of the theory comes from the constraints given by dynamical invariants. In the present paper, we will focus on the role played by the energy and the enstrophy constraints. In the framework of the MRS theory, it is justified in a low energy limit, and it is only a first step before more comprehensive studies including the effects of additional invariants, see e.g \cite{MajdaWangBook,CorvellecBouchet2010}.}

{The phenomenology of two-dimensional and geostrophic turbulence is often explained by considering the energy and enstrophy constraints only, see e.g \citet{Kraichnan_Motgommery_1980_Reports_Progress_Physics,Salmon_1998_Book}. Actually, earlier statistical mechanics approaches were developed for Galerkin-truncated Euler or quasi-geostrophic dynamics, which allowed for a simpler theoretical treatment than in the continuous dynamics case, see e.g. \cite{Kraichnan_1967PhFl...10.1417K,SalmonHollowayHendershott:1976_JFM_stat_mech_QG}. In the case of Galerkin-truncated models, only the energy and the enstrophy are conserved quantities, so the statistical mechanics of these systems were called energy-enstrophy theory.}

{A first reason to consider MRS theory rather than earlier energy-enstrophy statistical mechanics approach is that real flows are continuous, not Galerkin-truncated. Because real flows are also forced-dissipated, our first working hypothesis is that the computation of the MRS equilibrium states of the continuous dynamics is a necessary step before taking into account the effect of weak forcing and dissipation. This is for instance the approach followed by \cite{BouchetSimonnetPRL09}, who found that in the presence of weak forcing and dissipation, the dynamics were close to MRS equilibrium states which were different from energy-enstrophy equilibrium states.}

{One could  argue that numerical models often at best conserve energy and enstrophy, so there would be at first sight no reason to consider higher order invariants to predict the simulated self-organization. However, the truncated dynamics in numerical models may lead to the formation of quasi-stationary states having a life time that tends to infinity as the resolution is refined.  These quasi-stationary states are distinct from the energy-enstrophy equilibria, and thus require more invariants to describe using a statistical theory.  The formation of such quasi-stationary states is a characteristic of long-range interacting systems (e.g.,  \cite{CampaDauxoisRuffo} and references therein).  Given these results, our second working hypothesis is that the quasi-stationary states found in the truncated dynamics of numerical simulations may be close to a state predicted by MRS theory.}

{There are only a few studies of statistical equilibria of the continuously stratified quasi-geostrophic dynamics. \cite{Merryfield98JFM} obtained and discussed a class of solutions in the framework of energy-enstrophy statistical theory above topography, for which there is linear relation between potential vorticity $q$ and streamfunction $\psi$. The solutions obtained were critical points of the theory, but the actual entropy maxima were not selected. Here we consider the particular class of MRS equilibrium states characterized by a linear $q-\psi$ relation, so we describe the same class of equilibrium states, except that we do not take into account bottom topography. The effect of bottom topography will be addressed in a future work. The novelty of our work is to relate each critical point to the energy and fine-grained enstrophy constraints and to select the actual entropy maxima in different limit cases.}

\citet{SchecterPRE03} provided a direct comparison between numerical simulations
and numerical computation of MRS statistical equilibria, focusing on the
particular case of initial conditions given by a random field statistically
homogeneous in space, as in \citet{McWilliams94}. {This corresponds to a particular case for which the enstrophy profile is only weakly varying on the vertical.} We will outline in this paper the important consequence of the conservation of enstrophy at each depth. This important role played by the enstrophy constraint  allows us to discuss how the beta effect can favor barotropization. For a given initial streamfunction field, the beta effect modifies the initial distribution of potential vorticity, and therefore modifies the vertical profile of fine-grained enstrophy. More precisely, increasing the beta effect tends to increase the depth independent part of the fine-grained enstrophy profile. We will show that statistical equilibria associated with depth independent fine-grained enstrophy profiles are barotropic, which means that increasing the beta favors barotropization, according to statistical mechanics predictions.

The paper is organized as follows: the continuously stratified quasi-geostrophic equations are introduced on section \ref{sec:csqg}. The MRS equilibrium statistical mechanics theory of such systems is presented in section 3. Here it is shown that the class of MRS statistical equilibria characterized by a linear relation between potential vorticity and streamfunction are solutions of a minimum enstrophy problem, which is solved in various limit cases, first without beta effect and then with beta effect. The theory predicts an increase of barotropization with increasing value of beta, which is tested against numerical simulations in section 4. A summary of the main results and a discussion of their application to the ocean is given section 5.

\section{Continuously stratified quasi-geostrophic model} \label{sec:csqg}

\subsection{Equations of motion}

We consider an initial value problem for a freely evolving, unforced,
undissipated, quasi-geostrophic flow with continuous stratification
$N(z)$ (see e.g., \citet{VallisBook}, section 5.4):

\begin{equation}
\partial_{t}q+J(\psi,q)=0\ ,\label{eq:FullDynamics}\end{equation}
 \begin{equation}
q=\Delta\psi+\frac{\partial}{\partial z}\left(\frac{f_{0}^{2}}{N^{2}}\frac{\partial}{\partial z}\psi\right)+\beta y\ ,\label{eq:PV_definition}\end{equation}
 with $J(\psi,q)=\partial_{x}\psi\partial_{y}q-\partial_{y}\psi\partial_{x}q$
, and $f=f_{0}+\beta y$ is the Coriolis parameter with the $\beta$-plane
approximation.
Neglecting buoyancy and topography variations, the boundary condition at the bottom ($z=-H$, where $H$ is a constant) is given by 
\begin{equation}
 \partial_{z}\psi|_{z=-H}=0.	
\end{equation}
The boundary condition at the surface (defined as $z=0$,
using the rigid lid approximation), is given by the advection of buoyancy
\begin{equation}
 \partial_{t}b_{s}+J(\psi,b_{s})=0, \quad 
    \frac{f_{0}^{2}}{N^{2}}\partial_{z}\psi|_{z=0}=b_{s}.
\label{eq:FullDynamicsB}
\end{equation}

{The reason we focus on quasi-geostrophic dynamics  is that for these flows, there is  no  direct cascade of energy and no dissipative anomaly, see e.g. \cite{Salmon_1998_Book}, which makes possible a straightforward generalization of statistical mechanics theories developed initially in the framework of two-dimensional Euler equations. As far as the ocean is concerned, quasi-geostrophic dynamics is relevant to describe ``mesoscale turbulence'', but does not account for  ``sub-mesoscale turbulence'', which involve ageostrophic effects, with part of the energy that may cascade forward, see e.g. \cite{LeithThomas,FerrariWunsch09} and references therein.}

\subsection{Modal decomposition \label{sub:Vertical-and-horizontal-eigenmodes}}

\subsubsection{Laplacian and baroclinic eigenmodes}

We consider the case of a doubly periodic domain $\mathcal{D}=[-\pi\ \pi]\times[-\pi\ \pi]$.
The Fourier modes $e_{k,l}(x,y)$ are a complete orthonormal basis
of the 2D Laplacian $\Delta e_{k,l}=-K^{2}e_{k,l}$
where $(k,l)$ are the wavenumbers in each directions and $K^2 = k^2 + l^2$.
In the vertical, the complete orthonormal basis of barotropic-baroclinic
modes is defined by the solutions $F_{m}(z)$ of the Sturm-Liouville
eigenvalue problem 
\begin{equation}
\frac{\partial}{\partial z}\left(\frac{f_{0}^{2}}{N^{2}}\frac{\partial}{\partial z}F_{m}\right)=-\lambda_{m}^{2}F_{m}\quad\text{with}\quad F^{\prime}(0)=F^{\prime}(-H)=0, 
\label{eq:VerticalModes}
\end{equation}
 where $m\ge0$ is an integer, $\lambda_{m}^{2}$ is the eigenvalue
associated with mode $m$, which defines the $m$$^{th}$-baroclinic
Rossby radius of deformation $\lambda_{m}^{-1}$. The barotropic mode
is the (depth independent) mode $F_{0}$, associated with $\lambda_{0}=0$.

\subsubsection{Surface quasi-geostrophic (SQG) modes \label{sub:SQG_modes}}

In the following, it will be convenient to distinguish the interior
dynamics from the surface dynamics. For a given potential vorticity
field $q(x,y,z,t)$ and surface buoyancy field $b_{s}(x,y,t)$ , the
streamfunction can be written $\psi=\psi_\text{int}+\psi_\text{surf}$,  where $\psi_\text{int}$ is the solution of 
\begin{equation}
  \Delta\psi+\frac{\partial}{\partial z} 
  \left(\frac{f_{0}^{2}}{N^{2}}\frac{\partial}{\partial z}\psi\right)=q-\beta y\quad\text{with}\quad\partial_{z}\psi|_{z=0}=\partial_{z}\psi|_{z=-H}=0,
\label{eq:PV_interior}
\end{equation}
 and where $\psi_\text{surf}$ is the solution of 
\begin{equation}
\Delta\psi+\frac{\partial}{\partial z} \left(\frac{f_{0}^{2}}{N^{2}}\frac{\partial}{\partial z}\psi\right)=0, \text{~with ~~} \partial_{z}\psi|_{z=0}=\left(\frac{N(0)}{f_{0}}\right)^2 b_{s},
 \quad\partial_{z}\psi|_{z=-H}=0 \ .\label{eq:PV_surface}
\end{equation}

Each projection of $\psi_\text{surf}$ on a (horizontal) Laplacian eigenmode
$(l,k)$ defines a surface quasi-geostrophic eigenmode (SQG mode hereafter). When the stratification $N$ is constant and when $f_{0}/(NK)\ll H$, these SQG modes are decreasing exponential functions 
\begin{equation}
\widehat{\psi}_{\text{surf},k,l}=A_{k,l}e^{zNK/f_{0}} + O\left(\frac{h_{K}}{H}\right),
\label{eq:SQGmode_ExpDecay}
\end{equation}
where the coefficients $A_{k,l}$ are determined using the boundary
condition at the surface. When the stratification is not constant,
but when $f_{0}/N(0)K\ll H,$ the SQG modes are still characterized
by a decreasing exponential, with an $e$-folding depth $h_{K}=f_{0}/KN(0)$.

The distinction between surface and interior streamfunction is sometimes
useful to describe the dynamics , see e.g. \citet{LapeyreKlein06},
but a shortcoming of such a decomposition is that $\psi_\text{int}$ and
$\psi_\text{surf}$ are not orthogonal.

\subsection{Delta-sheet approximation and SQG-like modes \label{sub:SQGlike_modes}}

The boundary condition (\ref{eq:FullDynamicsB}) can be formally replaced by the condition of no buoyancy variation ($ \partial_z \psi = 0 $ at $z=0$), provided that surface buoyancy anomalies are interpreted as a thin sheet of potential vorticity just below the rigid lid \citep{Bretherton66}. For this reason, and without loss of generality, we will consider that $b_s=0$ {in the remainder of this paper}.  

Let us now  consider a case with no surface buoyancy, but with a surface intensified potential vorticity field defined as $q=q_\text{surf}(x,y)\Theta(z+H_{1})+q_\text{int}(x,y,z)$, with $H_{1}q_\text{surf}\gg Hq_\text{int}$, $H_{1}\ll H$ and $\Theta$ the Heaviside function. Using the linearity of (\ref{eq:PV_interior}), one can still decompose the streamfunction field into $\psi=\psi_\text{surf}+\psi_\text{int}$ where $\psi_\text{surf}$ is the flow induced by the PV field $q=q_\text{surf}\Theta(z+H_{1})$, obtained by inverting (\ref{eq:PV_definition}), and $\psi_\text{int}$ the flow induced by $q_\text{int}$. For $z<-H_{1}$, the streamfunction $\psi_\text{surf}$ satisfies equation (\ref{eq:PV_surface}), just as SQG modes. This is why we call it an SQG-like mode.

\section{Equilibrium statistical mechanics of stratified quasi-geostrophic
flow}\label{sec:Equilibrium-statistical-mechanics}

In this section we briefly introduce the MRS equilibrium statistical mechanics theory, and
introduce a general method to compute the MRS equilibria. This allows for explicit computation
of statistical equilibria on the $f-$plane in subsection
\ref{sub:Computation-of-the-RSM-f-plane}, and on the $\beta$-plane in subsection
\ref{sub:Including-the-beta}.

\subsection{Theory\label{sub:Theory}}

Let us consider an initial (fine-grained) PV field $q_{0}(x,y,z)$.
The dynamics (\ref{eq:FullDynamics}) induced by this PV field stirs
the field itself, which rapidly leads to its filamentation at smaller
and smaller scale, with the concomitant effect of stretching and folding.
There are two (related) consequences, one in phase space, one in physical
space.

In physical space, the filamentation toward small scales implies that
after a sufficiently long time evolution, the PV field can be described
at a macroscopic level by a probability field
$\rho(x,y,z,\sigma,t)\mathrm{d}\sigma$ to measure a given PV level
between $\sigma$ and $\sigma+\mathrm{d}\sigma$ at a given point
($x,y,z$). This probability distribution function (pdf hereafter)
$\rho$ is normalized locally \begin{equation}
\forall ~ x,y,z\ \mathcal{N}[\rho]=\int_{\Sigma}\mathrm{d}\sigma\ \rho=1,\label{eq:normalization}\end{equation}
where the integral is performed over the range $\Sigma$ of PV levels,
which is prescribed by the initial condition. The field $\rho$ is
a macrostate because it is associated with a huge number of fine-grained
PV configurations $q(x,y,z)$ (the microscopic states), in
the sense that many realizations of the (microscopic, or fine-grained)
PV field $q$ leads to the same macroscopic state $\rho$. 

The phase space is defined by all configurations of the field $q(x,y,z)$.
The cornerstone of the theory is to assume that  stirring
allows the system to explore evenly each possible configuration of
the phase space that satisfies the constraints of the dynamics, so
that these configurations can be considered equiprobable. When this assumption fails, as in the linear regime discussed in section 4.4, the statistical theory fails. 

The main idea of the theory is then to count the number of microstate
configurations associated with a macrostate (the pdf field $\rho$),
and to say that the equilibrium state is the most probable macrostate
that satisfies the constraints of the dynamics, i.e. the one that
is associated with the largest number of microstates satisfying the
constraints. One can then show that an overwhelming number of microstates
are associated with this most probable macrostate \citep{Michel_Robert_LargeDeviations94C}. An important physical consequence is that an observer who wants to see the actual equilibrium macrostate would neither need to perform ensemble average nor time average, but would simply have to wait a sufficiently long time.

Note that the output of the theory is the pdf field $\rho$, but the
quantity of interest to describe the flow structure is the coarse-grained PV field defined as 
\begin{equation}
\overline{q}=\int_{\Sigma}\mathrm{d}\sigma\ \sigma\rho.
\label{eq:q_coarse_grained}
\end{equation}
In practice, counting the number of microstates $q$ associated with
a macrostate $\rho$ is a difficult task, because $q$ is a continuous
field, and it requires the use of large deviation theory{, see e.g. \cite{BouchetCorvellec10}}. However,
it can be shown that classical counting arguments 
do apply here, and that the most probable state $\rho$ maximizes
the usual Boltzmann mixing entropy
\begin{equation}
\mathcal{S}=-\int_{\mathcal{D}}\mathrm{d}x\mathrm{d}y\ \int_{-H}^{0}\mathrm{d}z\ \int_{\Sigma}\mathrm{d}\sigma\ \rho\ln\rho\ ,
\label{eq:entropy_mixing}
\end{equation}
see  \citet{BouchetCorvellec10} and references therein for further details.

Let us now collect the constraints that $\rho$ must satisfy, in addition
to the local normalization (\ref{eq:normalization}). 
An infinite number of constraints are given by conservation
of the global distribution of PV levels $\mathcal{P}(\sigma,z)$, which is prescribed  by the initial global distribution of PV levels $P_{0}(\sigma,z)$: 
\begin{equation}
\mathcal{P}(\sigma,z)[\rho]=\int_{\mathcal{D}}\mathrm{d}x\mathrm{d}y\ \rho=P_{0}(\sigma,z).\label{eq:PV_constraint}\end{equation}
These conservation laws include the conservation of the fine-grained enstrophy
\begin{equation}
\int_{\mathcal{D}} \mathrm{d} x  \mathrm{d} y \int_{\Sigma} \mathrm{d} \sigma\ \sigma^2 \rho  =Z_0 \ . \label{eq:Z0} 
\end{equation}
Another constraint is given by the conservation of the total energy
  \begin{equation}
\mathcal{E}=\frac{1}{2}\int_{-H}^{0}\mathrm{d}z \int_{\mathcal{D}}\mathrm{d}x\mathrm{d}y\ \left(\left(\nabla\psi\right)^{2}+\frac{f_{0}^{2}}{N^{2}}\left(\partial_{z}\psi\right)^{2}\right) \ .
\label{eq:Energie_definition}\end{equation}
Because an overwhelming number of microstates are close to the equilibrium state $\rho$, the energy of the fluctuations are negligible with respect to the energy due to the coarse-grained PV field $\overline{q}$ \citep{Michel_Robert_LargeDeviations94C}. Integrating \ref{eq:Energie_definition} by parts and replacing $q$ by $\overline{q}$ gives the constraint  $E_{0}=\mathcal{E}[\rho]$, where $E_{0}$ is the energy of the initial condition, and where 
\begin{equation}
\mathcal{E}=\frac{1}{2} \int_{-H}^{0}\mathrm{d}z \int_{\mathcal{D}}\mathrm{d}x\mathrm{d}y \left(\beta y-\overline{q}  \right)\psi=\frac{1}{2} \int_{-H}^{0}\mathrm{d}z\int_{\mathcal{D}}\mathrm{d}x\mathrm{d}y \int_{\Sigma}\mathrm{d} \sigma  \left( \beta y-\sigma \right)\psi\rho\ .\label{eq:energy_functional_rho} 
\end{equation}
Finally, the MRS theory provides a variational problem
\begin{equation}
S\left(E_{0},P_{0}(\sigma,z)\right)=\max_{\rho,\mathcal{N}[\rho]=1}\left\{ \mathcal{S}\left[\rho\right]\ |\ \mathcal{E}\left[\rho\right]=E_{0}\ \&\ \mathcal{P}\left[\rho\right]=P_{0}(\sigma,z)\right\} \,,\label{eq:RSM}\end{equation}
which means that the equilibrium state is the density probability field
$\rho_\text{rsm}$ that satisfies the local normalization constraint (\ref{eq:normalization}), the energy constraint $\mathcal{E}\left[\rho\right]=E_{0}$ with $\mathcal{E}$ given by (\ref{eq:energy_functional_rho}),  the incompressibility constraint $\mathcal{P}(\sigma,z)\left[\rho\right]=P_{0}(\sigma,z)$ with $\mathcal{P}$ given by (\ref{eq:PV_constraint}), and that maximizes the entropy functional $\mathcal{S}$ given by (\ref{eq:entropy_mixing}). We see that the parameters of this problem are the values of the constraints, which are prescribed by the initial condition $q_{0}(x,y,z)$.

\subsection{MRS equilibrium states characterized by a linear $q-\psi$ relation \label{sub:Energy-enstrophy-equilibrium-states}}

\subsubsection{Simplification of the variational problem \label{sub:Simplification-of-the}}

In practice, MRS equilibrium states are difficult to compute, because
such computations involve the resolution of a variational problem
with an infinite number of constraints. Computing the critical points
of (\ref{eq:RSM}) is straightforward, see e.g. equation (\ref{eq:rhoChar})
of appendix A, but showing that second order variations of the entropy
functional are negative is in general difficult. However, recent results
have allowed considerable simplifications of the analytical computations
of these equilibrium states \citep{Bouchet:2008_Physica_D}.
The idea is first to prove equivalence (usually
for a restricted range of parameters) between the complicated variational
problem (\ref{eq:RSM}) and other variational problems more simple
to handle, involving less constraints than the complicated problem,
but characterized by the same critical points, and then to explicitly
compute solutions of the simpler variational problem.  The equivalence
between the variational problems ensures that solution of the
simple problem are also solutions of the more complicated one.
We will follow this approach in the following. 

\subsubsection{Minimum enstrophy variational problem \label{sub:MRS-linear}}

Starting from the MRS variational (\ref{eq:RSM}), it is shown in appendix A that any MRS equilibrium state characterized by a linear $\overline{q}-\psi$ relation is a solution of the variational problem
\begin{equation}
Z_\text{cg}^\text{tot}\left(E_{0},Z_{0}(z)\right)=\min_{\overline{q}}\left\{ \mathcal{Z}_\text{cg}^\text{tot}\left[\overline{q}\right]\ |\ \mathcal{E}\left[\overline{q}\right]=E_{0}\right\} \label{eq:MinimumEnstrophyVP}\end{equation}
with
\begin{equation}
\mathcal{Z}_\text{cg}^\text{tot}\left[\overline{q}\right]=\int_{-H}^{0}\mathrm{d}z\ \frac{\mathcal{Z}_{cg}}{Z_{0}},\quad\mathcal{Z}_{cg}=\frac{1}{2}\int_{\mathcal{D}}\mathrm{d}x\mathrm{d}y\ \overline{q}^{2},\label{eq:Enstrophy_coarse-grained}
\end{equation}
where $Z_{0}(z)$ is the fine-grained enstrophy profile, $E_{0}$
the energy. This variational problem amounts to find the minimizer
$\overline{q}_{m}$ (a coarse grained PV field) of the coarse-grained
enstrophy $\mathcal{Z}_{cg}^{tot}$ (with $\mathcal{Z}_{cg}^{tot}\left[\overline{q}_{m}\right]=Z_{cg}^{tot}\left(E_{0},Z_{0}(z)\right)$), among all the fields $\overline{q}$ satisfying the constraint $\mathcal{E}[\overline{q}]=E_{0}$ given by (\ref{eq:Energie_definition}).%

Critical points of the variational problem (\ref{eq:MinimumEnstrophyVP})
are computed by introducing a Lagrange multiplier $\beta_{t}$ associated
with the energy constraint, and by solving\[
\delta\mathcal{Z}_{cg}^{tot}+\beta_{t}\delta\mathcal{E}=0,\]
 where first variations of the functionals are taken with respect
to the field $\overline{q}$. Because $\beta_{t}$ is the Lagrange
parameter associated with the energy constraint, it is called inverse
temperature%
\footnote{The notation $\beta$ is traditionally used in thermodynamics for
inverse temperature. We use here the convention $\beta_{t}$ for this
inverse temperature, since $\beta=\partial_{y}f$ refers to the beta
effect, i.e. to the variation of the Coriolis parameter with latitude. %
}. Using $\delta\mathcal{Z}_{cg}^{tot}=\int_{\mathcal{D}}\mathrm{d}x\mathrm{d}y\ \int_{-H}^{0}\mathrm{d}z\ \overline{q}\delta\overline{q}/Z_{0}$
and $\delta\mathcal{E}=-\int\mathrm{d}x\mathrm{d}ydz\ \psi\delta\overline{q}$,
one finds that critical points of this problem are \begin{equation}
\overline{q}=\beta_{t}Z_{0}\psi.\label{eq:q-psi_relation_with_coeff_determined}\end{equation}
 The flow structure of these critical points can then be computed
by solving

\begin{equation}
\Delta\psi+\frac{\partial}{\partial z}\frac{f_{0}^{2}}{N^{2}}\frac{\partial}{\partial z}\psi+\beta y=Z_{0}\beta_{t}\psi\quad\text{with}\quad\partial_{z}\psi|_{z=0,z=-H}=0.\label{eq:CriticalPoints}\end{equation}
Critical points can then be classified according to the value of their
inverse temperature $\beta_{t}$. 
All the problem is then to find which of these critical point are
actual minima of the coarse-grained enstrophy $\mathcal{Z}_{cg}^{tot}[\overline{q}]$ for a given energy $\mathcal{E}[\overline{q}]=E_{0}$. %

MRS equilibrium states characterized by linear $\overline{q}-\psi$
relations are expected either in a low energy limit or in a strong
mixing limit, see \citet{BouchetVenaillePhysRep11} and references
therein. 
Let us estimate when the low energy limit is valid. For a given distribution
$P_{0}(\sigma,z)$ of PV levels $\sigma$, one can compute the maximum
energy $E_{max}$ among all the energies of the PV fields characterized
by the distribution $P_{0}(\sigma,z)$. Low energy states characterized
by the same distribution $P_{0}(\sigma,z)$ are therefore those characterized
by $E\ll E_{max}$. 
The strong mixing limit, which can in some case overlap the low energy
limit, is obtained by assuming $\beta_{t}\sigma\psi\ll1$ in the expression
(\ref{eq:rhoChar}) of the critical points of the MRS variational
problem (\ref{eq:RSM}), see \citep{ChavanisSommeria:1996_JFM_Classification}.  {Note that these results would remain valid in the presence of bottom topography. The only change would be the expression of the energy, and the bottom boundary condition.}

\subsubsection{Link with the phenomenological principle of enstrophy minimization and other equilibrium states}

At a phenomenological level, the variational problem (\ref{eq:MinimumEnstrophyVP}) can be interpreted as a generalization of the minimum enstrophy principle
of \citet{BrethertonHaidvogel} to the continuous stratified flow:
the equilibrium state minimizes the vertical integral of the coarse-grained
enstrophy normalized by the fine-grained enstrophy in each layer,
with a global constraint given by energy conservation. This normalization
is necessary to give a measure of the degree of mixing comparable
from one depth $z$ to another. 

It has been shown by \cite{Carnevale_Frederiksen_NLstab_statmech_topog_1987JFM} that solutions of the minimum enstrophy problem of \cite{BrethertonHaidvogel} for one layer quasi-geostrophic flows, are energy-enstrophy equilibrium states of the Galerkin-truncated dynamics of these models, in the limit when the high-wave number cut-off goes to infinity. These energy-enstrophy equilibrium states were first introduced in the framework of 2D truncated Euler flows by \cite{Kraichnan_1967PhFl...10.1417K}. \cite{SalmonHollowayHendershott:1976_JFM_stat_mech_QG} generalized the theory to one or few layers truncated quasi-geostrophic flows. \cite{FrederiksenSawford} applied the theory for a barotropic flow on a sphere, and \cite{fredericksen91GAFDa} generalized these results for two-layers quasi-geostrophic flow on a sphere. In this latter case, interesting stability results were obtained by \cite{FrederiksenB}.

The case of continuously stratified flow above topography was addressed by \citet{Merryfield98JFM} for truncated dynamics. \citet{Merryfield98JFM} computed critical points of the statistical mechanics variational problem, found $\overline{q}=\mu_t(z) \psi$ and discussed the vertical structures of these flows depending on $\mu_t$. We showed in the previous subsection that the parameter $\mu_t$ is related to the fine-grained enstrophy profile through $\mu_t=\beta_t Z_0(z)$. In addition, the critical states described by  \citet{Merryfield98JFM} could either be saddle, minima or maxima of the entropy functional. We show in the next section that we can select which of them are actual entropy maxima (or equivalently coarse-grained enstrophy minima), in different limit cases.

\subsubsection{Transients, mean and sharp equilibrium states}

{In the framework of  energy-enstrophy theory, a distinction is often made between ``transients'' and ``mean'' energy, see e.g. \cite{SalmonHollowayHendershott:1976_JFM_stat_mech_QG,fredericksen91GAFDa,MajdaWangBook}, among others. Statistical equilibria are called ``sharp'' when the ``transients'' vanish. In earlier works, the ``mean'' quantities were often computed by summing over all the microstates distributed according to the canonical measure, which were implicitly assumed to be equivalent to the microcanonical measure.  In the absence of topography and beta plane, the ``mean'' states defined in this way are always zero in a doubly periodic domain, whatever the resolution of the truncated model, and all the energy is in the ``transients''.}

{By contrast, the equilibrium states computed in the MRS framework in the \textit{microcanonical ensemble} are always sharp: the energy of the fluctuations necessarily vanish, which can be shown with large deviation theory, see \cite{BouchetCorvellec10} and references therein. This apparent paradox is solved when one realizes that microcanonical and canonical ensembles are, in general, not equivalent for long-ranged interacting systems, see e.g. \cite{Ellis00} for general considerations and \cite{VenailleBouchetJSP11} for application to two-dimensional and geophysical flows.} 

{In the case of a doubly periodic domain without bottom topography and beta effect, the MRS equilibrium states (or the energy-enstrophy equilibria) are degenerate, because of the symmetries of the domain, see \citep{BouchetVenaillePhysRep11} for more details and for a discussion on how to remove the degeneracy, see also \cite{Corentin} for the case of a spherical domain. In practice, the dynamics breaks the system symmetry by selecting one of the degenerate equilibrium states. Of course, if one considers the average potential vorticity field over all the degenerate equilibria, then one recovers zero, and this is what happens when computing the ``mean'' state in the canonical ensemble. We stress that for either quasi-geostrophic flows or  truncated quasi-geostrophic flows in the limit of infinite resolution, the freely evolving dynamics itself can not jump from one state to another once the symmetry is broken: in the microcanonical ensemble, there are no ``transients''. We also stress that the physically relevant statistical ensemble for an isolated system such as a freely evolving flow is the microcanonical ensemble.}

\subsection{Enstrophy minima on a $f$-plane \label{sub:Computation-of-the-RSM-f-plane}}

In the previous subsection, we have shown that MRS equilibrium states
characterized by a linear $\overline{q}-\psi$ relationship are solution
of a minimum coarse-grained enstrophy variational problem (\ref{eq:MinimumEnstrophyVP}). Such MRS equilibrium states will be referred to as ``coarse-grained
enstrophy minima'' in the following. We found in the previous section
that the affine relation of these coarse-grained enstrophy minima
are of the form $\overline{q}=\beta_{t}Z_{0}\psi$, with
a coefficient proportional to the fine-grained enstrophy $Z_{0}$,
and are solutions of (\ref{eq:CriticalPoints}). The critical points
of the variational problem (\ref{eq:MinimumEnstrophyVP}) are any
states $\overline{q}$ satisfying these conditions. The aim of this
subsection is to find which of these critical point $\overline{q}$
are the actual coarse-grained enstrophy minima, solutions of variational
problem (\ref{eq:MinimumEnstrophyVP}), when there is no $\beta-$effect,
but for arbitrary stratification $N$.

Injecting (\ref{eq:q-psi_relation_with_coeff_determined}) in (\ref{eq:Enstrophy_coarse-grained}),
and using the expression (\ref{eq:energy_functional_rho}) with the
constraint $E_{0}=\mathcal{E}[\rho]$, one finds that the coarse-grained
enstrophy of each critical point $\overline{q}$ is given by \begin{equation}
\mathcal{Z}_\text{cg}^\text{tot}=-\beta_{t}E_{0}\ .\label{eq:enstrophy_critical_point_simple_case}
\end{equation}
 Remarkably, the coarse grained enstrophy of a critical point $\overline{q}$
depends only on the inverse temperature $\beta_{t}$ . We conclude
that coarse-grained enstrophy minimum states are the solutions of
(\ref{eq:CriticalPoints}) associated with the largest value $\beta_{t}$.

\begin{figure}
\begin{center}
\includegraphics[width=\textwidth]{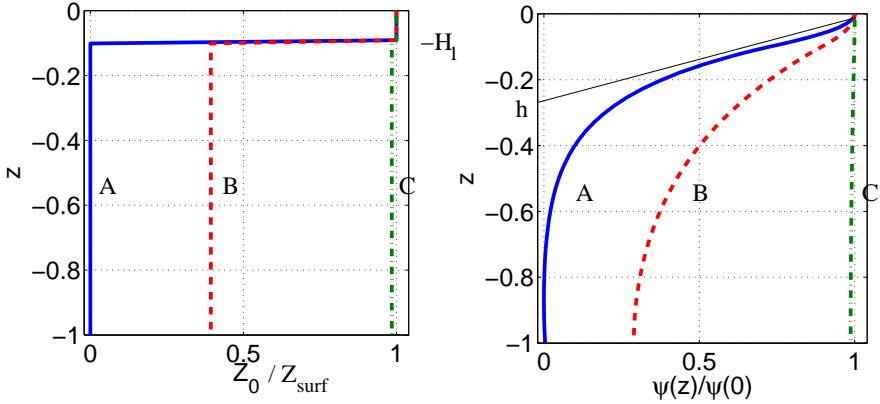}
\caption{Left panel: three different fine grained enstrophy profiles. Right
panel: corresponding vertical structure of statistical equilibrium
states ($\psi(z)/\psi(0)$ on the left panel), in the case of constant
stratification ( $f_{0}^{2}/N^{2}=0.1$). The $e$-folding depth in
case A is $h=f_{0}/NK$ (with here $K=1$ for the statistical equilibrium
state). 
\label{fig:e-folding_depth}}
\end{center}
\end{figure}

Projecting (\ref{eq:CriticalPoints}) on the Laplacian eigenmode
$e_{l,k}(x,y)$ gives \begin{equation}
\frac{\partial}{\partial z}\left(\frac{f_{0}^{2}}{N^{2}}\frac{\partial}{\partial z}\widehat{\psi}_{k,l}\right)=\left(\beta_{t}Z_{0}+K^{2}\right)\widehat{\psi}_{k,l}\ ,
\label{eq:VertStruc}
\end{equation}
with
\[
\partial_{z}\widehat{\psi}_{k,l}|_{z=0,-H}=0,\quad K^{2}=k^{2}+l^{2},\quad \psi=\sum_{k,l}\widehat{\psi}_{k,l}e_{kl}.\]
We see that each critical point is characterized by a given wavenumber modulus $K$. Its vertical structure and the corresponding value of $\beta_t$ must be computed numerically in the case of arbitrary profiles $Z_0(z)$. 
Let us  consider the example shown in Fig.\ \ref{fig:e-folding_depth}, for a two-step fine-grained enstrophy profile 
\begin{equation}
Z_{0}=Z_\text{surf}\Theta\left(z+H_{1}\right)+Z_\text{int}\Theta\left(-z-H_{1}\right),\quad H_{1}\ll H,\label{eq:Z0_profile_PieceWise_TwoLayers}
\end{equation}
where $\Theta$ is the Heaviside function, and for $Z_\text{int}/Z_\text{surf}$
varying between $0$ and $1$. 
We find that the minimum coarse-grained
enstrophy states are always characterized by the gravest horizontal
mode on the horizontal ($K=1$). As for the vertical structure, we
observe Fig.\ \ref{fig:e-folding_depth} a tendency toward more barotropic
flows when the ratio $Z_\text{int}/Z_\text{surf}$ tends to one.

It is instructive to consider the limit cases $Z_\text{surf}=Z_\text{int}$
and $Z_\text{int}=0$, corresponding respectively to case A and C of Fig.\ \ref{fig:e-folding_depth},
for which minimum enstrophy can be explicitly solved.
When the enstrophy profile is depth independent ($Z_{0}(z)=Z_\text{int}$,
case C of figure \ref{fig:e-folding_depth}), solutions of (\ref{eq:VertStruc})
are given by the modes $F_{m}(z)e_{K}(x,y)$ defined in subsection
(\ref{sub:Vertical-and-horizontal-eigenmodes}), and are associated
with inverse temperatures $\beta_{t}=-\left(\lambda_{m}^{2}+K^{2}\right)/Z_\text{int}$. We see that the maximum value of $\beta_{t}$ is reached for the gravest
horizontal mode ($K=1$) and the gravest vertical mode $(\mbox{\ensuremath{\lambda}}_{0}=0)$.
It means that the coarse-grained enstrophy minimum state is barotropic.

When $Z_\text{int}=0$ (case A of Fig. \ref{fig:e-folding_depth}), it
is straightforward to show that critical points, the solutions of
(\ref{eq:VertStruc}), are SQG-like modes (see subsection \ref{sub:SQGlike_modes})
of $e$-folding depth $h=f_{0}/N(0)K$, associated with $\beta_{t}=-N(0)K/(H_{1}Z_\text{surf}f_{0})$
with corrections of order $H_{1}/H$ and of order $h/H$. The coarse-grained
enstrophy minimum is therefore the SQG-like mode associated with $K=1$
and with the largest $e-$folding depth $h=Lf/2\pi N$, where $L$
is the domain length scale. 

These examples show the importance of the conservation of fine-grained
enstrophy to the vertical structure of the equilibrium state. The
main result is that statistical mechanics predicts a tendency for
the flow to reach the gravest Laplacian mode on the horizontal. The
vertical structure associated with this state is fully prescribed
by solving (\ref{eq:VertStruc}) with $K=1$. Because the barotropic
component of such flows are larger than solutions of (\ref{eq:VertStruc})
with $K>1$, we can say that the inverse cascade on the horizontal
is associated with a tendency to reach the gravest vertical mode compatible
with the vertical fine-grained enstrophy profile $Z_{0}$. This means
a tendency toward barotropization, although in general, the fact that
the profile $Z_{0}$ is non constant prevents complete barotropization.%

{Note that the previous results are presented in the case of a doubly periodic domain, but it would be straightforward to generalize them to channel geometries, or to any other domains with boundaries.  The only change would be the spatial structure of the Laplacian eigenmodes. For these domain geometries, one would also  need to discuss the effect of the circulation values at the boundaries.}

\subsection{Including the $\beta$-effect \label{sub:Including-the-beta}.}

We now discuss how the results from the previous section generalize in the presence of beta. For a given initial condition $\psi_{0}(x,y,z)$, increasing $\beta$ increases the contribution of the (depth independent) available potential enstrophy defined as 
\begin{equation}
Z_{p}=\beta^{2}\int_{\mathcal{D}}\mathrm{d}x\mathrm{d}y\ y^{2},\label{eq:Potential_Enstrophy}
\end{equation}
to the total fine-grained enstrophy profile $Z_{0}(z)=\int_{\mathcal{D}}\mathrm{d}x\mathrm{d}y\ q_{0}^{2}$, where $q_{0}$ is the initial PV field that can be computed by injecting $\psi_{0}$ in (\ref{eq:PV_definition}). For sufficiently large values of $\beta$, the PV field is dominated by the beta effect ($q_0 \approx \beta y $),  $Z_{0}$ therefore tends to $Z_{p}$ and becomes depth independent. Because statistical equilibria computed in the previous subsection were fully barotropic when the fine-grained enstrophy $Z_{0}$ was depth-independent, we expect a tendency toward barotropization by increasing $\beta$.

Consider for instance the case where $\psi_{0}$ is an SQG-like mode
such that $q_{0}=\beta y$ for $z<-H_{1}$, with $H_{1}\ll H$ and
associated with a surface fine-grained enstrophy $Z_{0}=Z_\text{surf}$
for $z>-H_{1}$. The initial profile of fine-grained enstrophy $Z_{0}$
is thus given by (\ref{eq:Z0_profile_PieceWise_TwoLayers}), with
the interior enstrophy $Z_\text{int}=Z_{p}$ prescribed by the available
potential enstrophy (\ref{eq:Potential_Enstrophy}). The question
is then to determine whether the results obtained previously without
beta effect still hold in presence of it.

{In oder to apply properly MRS theory in the presence of beta, one must consider southern and northern boundaries. Indeed, MRS theory relies on the conservation of the global fine-grained PV distribution. Strictly speaking, this  distribution is not conserved in the doubly periodic configuration on a beta plane, since any fluid particle that travels $2\pi$ in the $y$ direction gains a value $-2 \pi \beta$, and one can not know a priori how many times a fluid particle will turn over the domain in the $y$ direction before the flow reaches equilibrium. In practice, the theory can  still be used in the doubly periodic case to make qualitative predictions, as explained in the next section.}

{In the channel configuration, the streamfunction is a constant at $y= \pm \pi$, and we can compute the statistical equilibria in two limit cases. First, when $\beta=0$, we recover the configuration with surface intensification of the fine grained enstrophy profile $Z_{0}$, for which we found that statistical equilibria were the SQG-like modes associated with the gravest Laplacian horizontal modes, see subsection \ref{sub:Computation-of-the-RSM-f-plane}.  Second, with $\beta$ sufficiently large, such that $Z_{p}\gg Z_\text{surf}$, the statistical equilibria can be computed by considering the case of a constant fine-grained enstrophy profile $Z_{0}$. In the general case, the values of the streamfunction at the northern and southern boundaries are different, and are determined using constraints given by mass conservation and  circulation conservation along each boundary, see e.g. \citep{PedloskyBook}. Here we consider the simple case  $\psi(x,y=\pm\pi,z)=0$, with zero total circulation ($\int_{\mathcal{D}}  \mathrm{d} x \mathrm{d} y q=0$) at each depth. It is necessary to take into account the conservation of linear momentum (associated with translational invariance of the problem in the $x$ direction), which provides an additional constraint $\mathcal{L}[\overline{q}]=L_{0}$, with}
\begin{equation}
\mathcal{L}[\overline{q}]= \int_{-H}^0 \mathrm{d} z \ \int_{\mathcal{D}} \mathrm{d}x\mathrm{d}y\ \overline{q}y.\label{eq:Momentum_y} \end{equation}
{Critical points of the variational problem (\ref{eq:MinimumEnstrophyVP}) with this additional constraint satisfy}
\begin{equation}
\overline{q}=\left(\beta_{t}  \psi -\mu y\right)Z_{0}, \label{eq:q-psi_with_linear_momentum}
\end{equation}
{where $\mu$ and $\beta_t$ are respectively the Lagrange multipliers associated with the linear momentum conservation and the energy conservation. Equilibrium states therefore satisfy} 
\begin{equation}
Z_{0}\left(\beta_{t} \psi - \mu y\right)=\Delta\psi+\frac{\partial}{\partial z}\left(\frac{f_{0}^{2}}{N^{2}}\frac{\partial}{\partial z}\psi \right) +\beta y\ . \label{eq:critical_points_channel}
\end{equation}
{Taking $\mu=-\beta/Z_{0}$, we recover the $f$-plane case of the previous subsection: solutions of Eq. (\ref{eq:critical_points_channel}) are barotropic-baroclinic modes on the vertical with eigenvalue $-\lambda_m^2$, and are Laplacian eigenmodes in the channel (with eigenvalue denoted by $-K^2$), with $\beta_t Z_0=-K^2+\lambda_m^2$. These states characterized by $\beta_{t}=-\left(K^2+\lambda_m^2\right)/Z_{0}$, $\mu=-\beta/Z_{0}$  are drifting towards the west at speed of  Rossby waves: $U_{drift}=-\beta/\left(\lambda_m^2+K^2\right)$. The enstrophy minimizer among these states is the one associated with the smallest value of $\beta_t$, namely the  barotropic mode associated  with the gravest Laplacian eigenmode consistent with the circulation and momentum constraints. When $\mu \ne \beta/Z_0$, Eq. (\ref{eq:critical_points_channel}) can be inverted, and $\psi$ is barotropic since $(\beta-\mu Z_0) y$ is depth-independent. The computation of the minimum enstrophy state is therefore equivalent to computing the minimum enstrophy state of a barotropic flow in a channel. A complete discussion of minimum enstrophy states in a barotropic channel are presented in \cite{CorvellecPhD}, see also \cite{MajdaWangBook}, with detailed computations including the effect of bottom topography and of non-zero circulations at the boundaries. The important point here is that in the large beta limit, the statistical equilibrium states are barotropic.}

\section{Numerical simulations \label{sec:Numerical-simulations}}

\subsection{Experimental set-up\label{sub:Numerical-settings}}

We consider in this section the final state organization of an initial SQG-like
mode (defined in subsection \ref{sub:SQGlike_modes}), varying the values of
$\beta$ in a doubly periodic square domain of size $L=2\pi$. More precisely, we
consider a vertical discretization and an horizontal Galerkin truncation of the
dynamics (\ref{eq:FullDynamics}), for an initial potential vorticity field
$q_{0}=q_{0,surf}(x,y)\Theta\left(z+H_{1}\right)+\beta y$, such that $q_{0}=\beta
y$ in the interior ($-H<z<-H_{1}$) and $q_{0}\approx q_{0surf}$ in a surface layer
$z>-H_{1}$. The surface PV $q_\text{surf}(x,y)$ is a random field with random
phases in spectral space, and a Gaussian power spectrum peaked at wavenumber
$K_{0}=5$, with variance $\delta K_{0}=2$, and normalized such that the total
energy is equal to one ($E_{0}=1$). This corresponds to the case discussed
subsection (\ref{sub:Including-the-beta}), with a vertical profile of fine-grained
enstrophy given by (\ref{eq:Z0_profile_PieceWise_TwoLayers}), where the interior
enstrophy is given by the available potential enstrophy
(\ref{eq:Potential_Enstrophy}): for $z<-H_{1}$, $Z_{0}=4\pi^{4}\beta^{2}/3$.

We perform simulations of the dynamics by considering a vertical discretization
with $10$ layers of equal depth, horizontal discretization of $512^{2}$, $H=1$,
and $F=(Lf_{0}/HN)^{2}=1$, using a pseudo-spectral quasi-geostrophic model for the horizontal dynamics
 without small scale dissipation \citep{SmithVallis01}. We choose $H_{1}=H/10$ for the initial condition, so that there is non zero enstrophy only in the upper layer in the absence of a beta
effect. We also perform experiments in the presence of small scale dissipation
(Laplacian viscosity or hyperviscosity), and found no qualitative differences as
far as the large scale horizontal structure and the vertical structure of the flow
were concerned. The only difference in presence of small scale dissipation is the
decay of the enstrophy in each layer, because small scale dissipation smooths out
PV filaments. Because of the energy cascade, the energy remains constant at lowest
order even in presence of small scale dissipation.

Time integration of the freely evolving dynamics proceeds for about $250$ eddy turnover times.
Typically, i) the unstable initial condition leads to a strong turbulent stirring and the flow
self-organizes rapidly into a few vortices, which takes a few eddy turnover times ii) same sign
vortices eventually merge together on a longer time scale (a few dozen of eddy turnover times),
and iii) the remaining dipole evolves very slowly on a slow time scale (more than hundreds of
eddy turnover times). In absence of $\beta$-effect, a good indicator of the convergence is given
by the convergence of the $q-\psi$ relation to a single functional relation. As we shall see in
the following, this phenomenology is complicated by the presence of beta.

{We saw in the previous section that a proper application of the MRS theory in presence of a beta plane would require considering northern or southern boundaries, since the fine-grained PV distribution is not conserved in the doubly-periodic domain case. However, we still can use the theory to interpret qualitatively the simulations in the doubly periodic case.  For small values of $\beta$, planetary vorticity gradients are negligible with respect to relative vorticity and stretching terms in the PV expression, in which case one can check that the fine grained PV distribution remains nearly the same during the flow evolution, and MRS can be applied. For large values of $\beta$, the fluid particles can not escape in the meridional direction, they are confined in zonal bands having a width of the order of the Rhines scale\footnote{Here $E_{0}=1$ is the initial energy, so that $E_{0}^{1/2}$ is a good metric for typical velocities of the flow (assuming it becomes barotropic).} $L_R = 2 \pi \sim E_0^{1/4}/\beta^{1/2}$, and thus the fine-grained PV distribution remains close to the initial one, and MRS theory can be applied.} 

\subsection{SQG-like dynamics ($\beta = 0$) \label{sub:SQG-like-dynamics-(small-beta)}}

\begin{figure}
\begin{center}
\includegraphics[width=\textwidth]{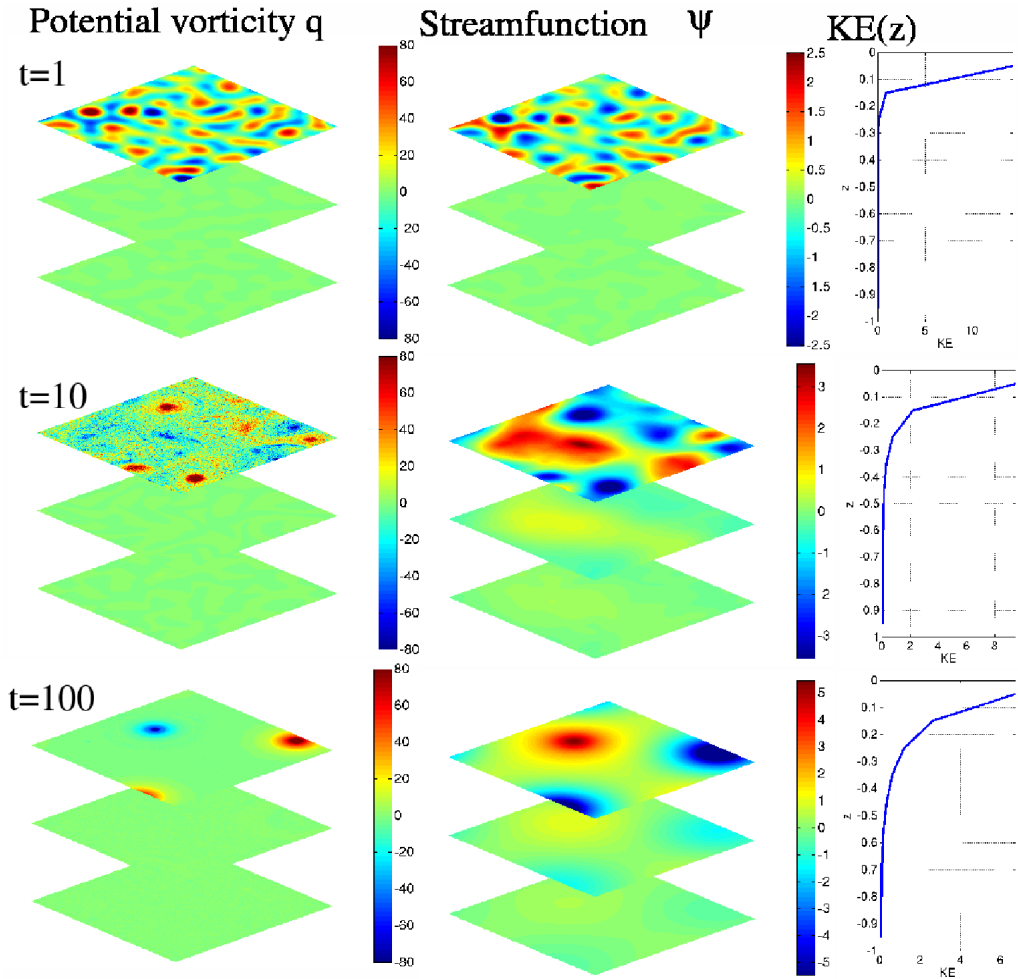}
\caption{case $\beta=0$ (SQG-like). Only the fields in upper, middle and lower
layer are shown.
 \label{fig:beta2}}
\end{center}
\end{figure}

The case with $\beta=0$ is presented in Fig. \ref{fig:beta2}.
The initial PV and streamfunction fields have horizontal structures
around wave number $K_{0}=5$, which is associated with an SQG-like
mode of typical e-folding depth $f_{0}/(NK_{0})\approx0.1$. 

An inverse cascade in the horizontal leads to flow structures with an horizontal wavenumber $K$ decreasing with time, associated with a tendency toward barotropization: the $e$-folding depth of the SQG-like mode increases as $f_{0}/2NK$. According to statistical mechanics arguments of the previous section, the flow should tend to the ``gravest'' horizontal SQG-like mode ($K=1$), which is also the ``gravest'' vertical SQG mode, i.e. the one associated with the largest $e$-folding depth. The concomitant horizontal inverse cascade (most of the kinetic energy is in the gravest horizontal mode $K=1$ at the end of the simulation) and the increase of the $e$-folding depth are observed on Fig. \ref{fig:beta2}, showing good qualitative agreement between statistical mechanics and numerical simulations.

\begin{figure}
\begin{center}
\includegraphics[width=0.5\textwidth]{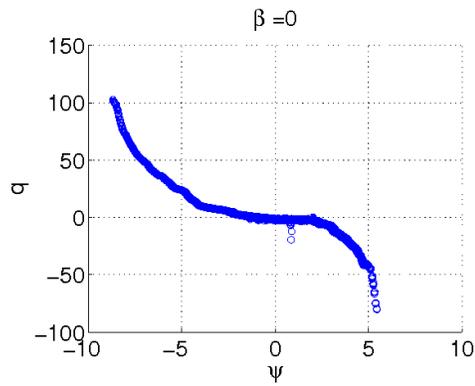}
\caption{$q-\psi$ relation in the upper layer for $t=250$ eddy turnover time.
\label{fig:q-psiBETA0}}
\end{center}
\end{figure}

 {There are however quantitative differences between the observed final state organization and the prediction of MRS equilibria characterized by a linear $\overline{q}-\psi$ relation. As seen on Fig. \ref{fig:q-psiBETA0}, the $\overline{q}-\psi$ relation is well defined, but it is not linear. Taking into account higher order invariants in the MRS theory would allow to better describe this $\sinh$-like relation, see e.g. \citep{BouchetSimonnetPRL09}, but it may not be sufficient to describe the ``unmixed'' cores of the remaining vortex in the final dipole \citep{SchecterPRE03}.}

The presence of these ``unmixed''  PV blobs implies the existence of structures smaller than wavenumber $K=1$. Because each projection of the PV field on a different wavenumber is associated with an SQG-like mode associated with a different $e$-folding depth, the vertical structure of the flow is actually a linear combination of SQG-like modes associated with different wavenumbers. A consequence is that the effective e-folding depth of the kinetic energy is $f_{0}/(2NK^\text{eff})$. We estimated the coefficient $K_\text{eff}\approx3.5$ by considering the linear $q-\psi$ relation passing through the extremal points of the observed $q-\psi$ relation of Fig. \ref{fig:q-psiBETA0}. {Let us conclude on this case by discussing the relaxation towards equilibrium: we observed that the $q-\psi$ relation did not change at all between $150$ and $250$ eddy turnover times, proving that the system reached a stationary state. In addition, we see in  Fig. \ref{fig:Emodes}-a that the kinetic energy of each vertical mode (and therefore of each layer) reaches a plateau after a few eddy turnover times. As expected from the MRS theory, the statistical equilibrium is sharp, there is no transient.}

\subsection{Effect of beta ($\beta \sim 1$)}

\begin{figure}
\begin{center}
\includegraphics[width=\textwidth]{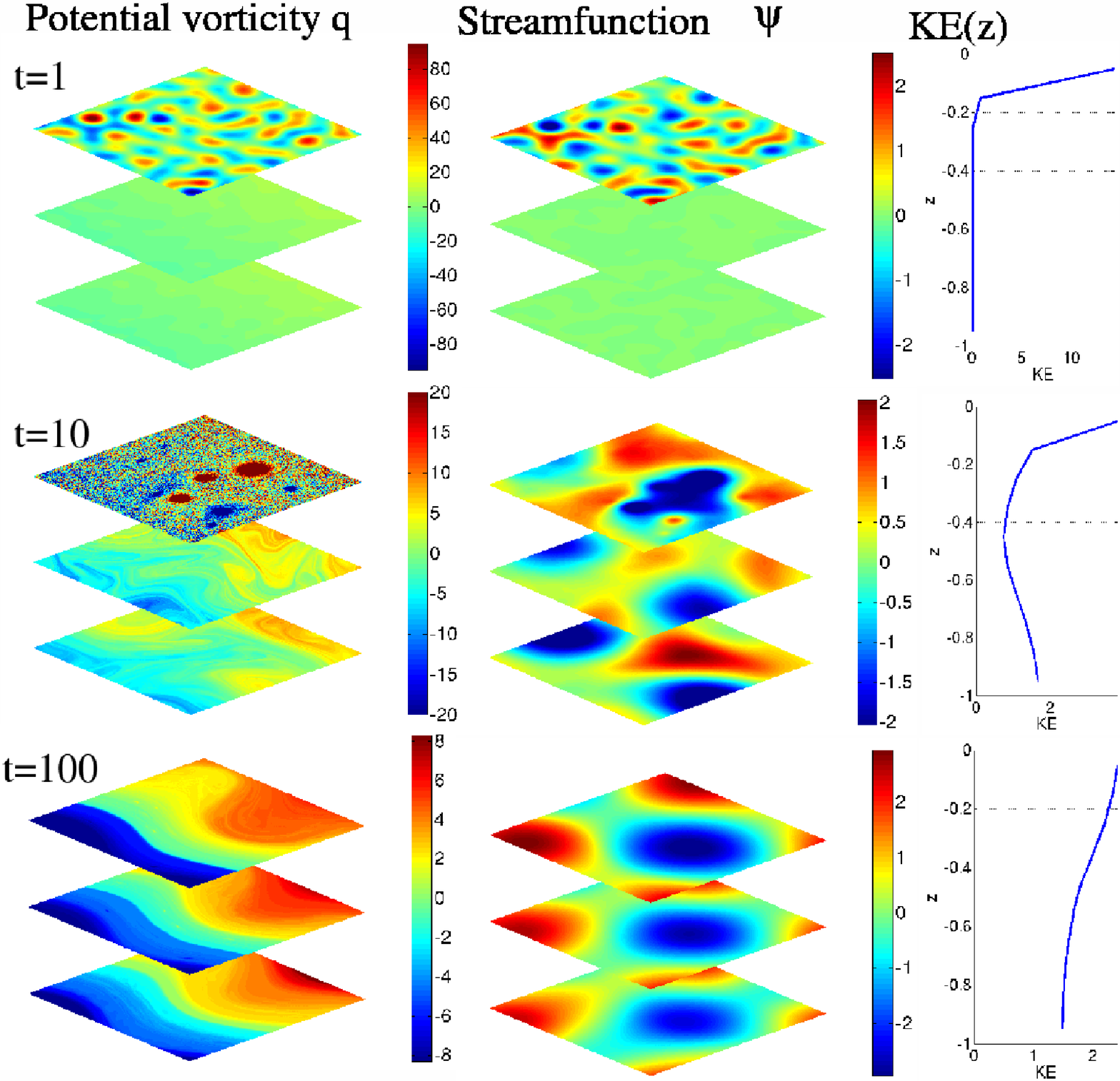}
\caption{case $\beta=2.1$. See Fig.\ \ref{fig:beta2} for legend. Note that
the colors scales are changed at each time; this is why the interior `` beta plane`` is not visible in the upper panel of potential
vorticity. 
\label{fig:beta3}}
\end{center}
\end{figure}

We now consider the free evolution of the same initial condition $\psi_{0}(x,y,z)$
as in the previous subsection, but on a beta plane. Because
$\beta\ne0$, the initial condition for potential vorticity is different
than in the previous section: the total energy and initial velocity
fields are the same as before, but the vertical profile of fine grained
enstrophy is different, since there is now available potential enstrophy
in the interior. As explained in subsection \ref{sub:Computation-of-the-RSM-f-plane}, this contribution from interior fine-grained enstrophy shall increase
the relative barotropic component of the statistical equilibrium state.
This is what is actually observed in the final state organization
of Fig. \ref{fig:beta3}. We conclude that in this regime, the beta
effect is a catalyst\footnote{By catalyst we do not mean a dynamical effect, as in chemistry, but simply the fact that barotropization in the final state is enhanced in presence of a beta plane.} of barotropization, as predicted by statistical mechanics.

\begin{figure}
\begin{center}
\includegraphics[width=0.5\textwidth]{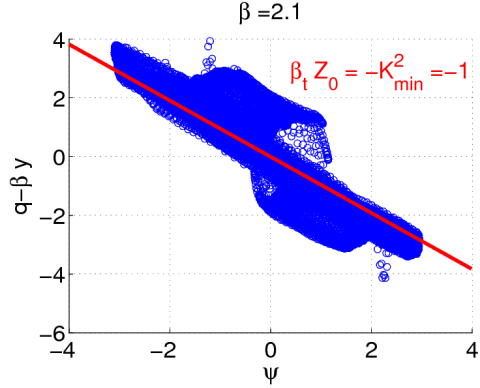}
\caption{{Relation between $q-\beta y$ and streamfunction in the upper layer, at $t=100$ eddy turnover times, for  $\beta=2.1$. The fact that this relation is close to be linear shows that the small energy limit considered in this paper is relevant to this case. The coefficient of the slope is approximatively one, meaning that the energy is condensed in the gravest horizontal Laplacian eigenmode, which is propagating westward at the speed of barotropic waves. This is consistent with the computation of minimum enstrophy states on a beta plane, see subsection \ref{sub:Including-the-beta}}. \label{fig:q-psiBETA1}}
\end{center}
\end{figure}

{Additionally, the $q-\psi$ relation presented in Fig. \ref{fig:q-psiBETA1} presents a very good agreement with the prediction of our computation of barotropic MRS equilibria carried in subsection \ref{sub:Including-the-beta}: we observe a linear relation between $(q-\beta y)$ and $\psi$ with a slope $\beta_{t}Z_{0}=-1$. This means that the dipole structure observed in the streamfunction field is the actual gravest horizontal Laplacian eigenmode $K=1$, and that this eigenmode is drifting westward at a speed $\beta/K^{2}$. In addition, we see in Fig. \ref{fig:Emodes} that after a few eddy turnover times, almost all the kinetic energy is in the barotropic mode, and that the energy levels in each vertical mode reach a plateau: there are no transients.}

{We conjecture that the good agreement with MRS theory occurs since the initial distribution of PV levels and the initial energy place the flow in the low energy regime considered in the Appendix. In addition, for  intermediate values of $\beta$, the Rhines scale $L_R=2\pi E_O^{1/4}/\beta^{1/2}$ is of the order of the domain scale. The beta effect limit the meridional displacements of fluid particles at this scale; this is why the computations performed Appendix 2 in a channel geometry are qualitatively relevant to this case.} 

\subsection{Effect of strong Rossby waves  ($\beta\gg1$) \label{sub:High-values-of-beta}}

\begin{figure}
\begin{center}
\includegraphics[width=\textwidth]{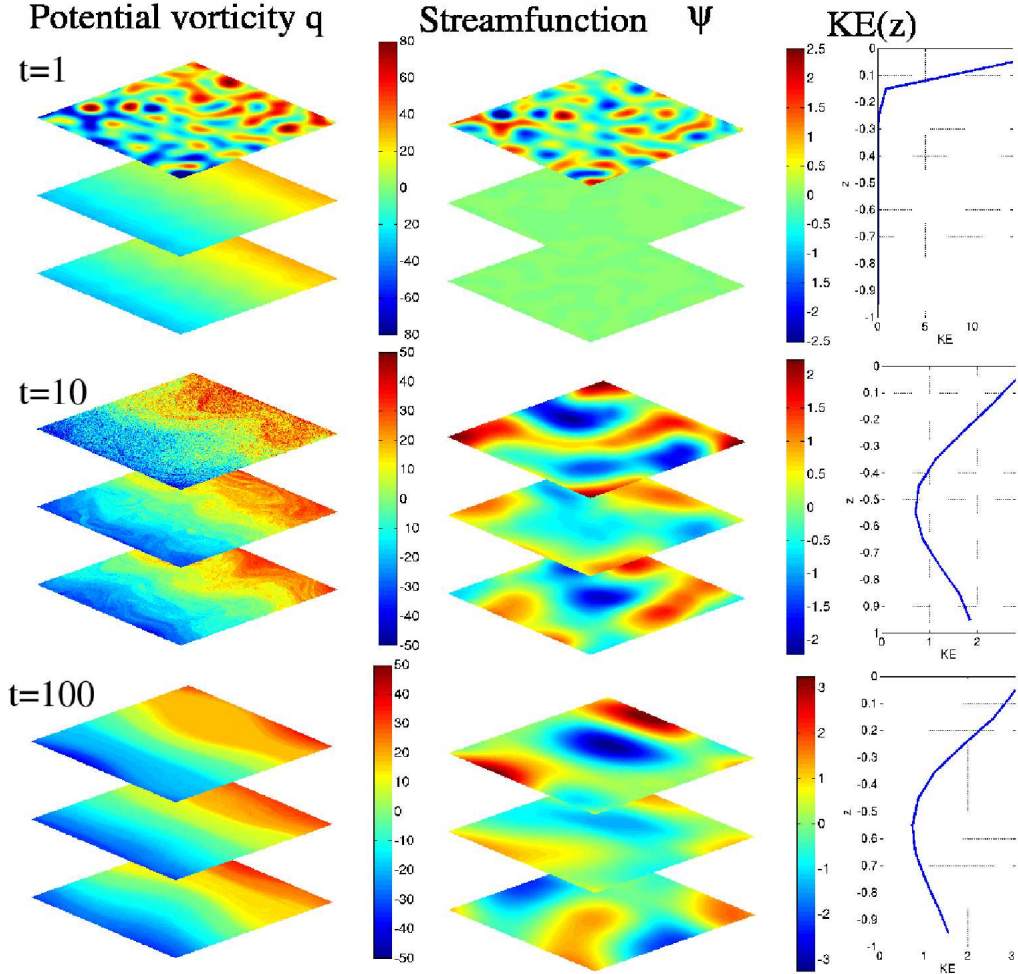}

\caption{case $\beta=10$, see Fig. \ref{fig:beta2} for legend. \label{fig:beta4} }
\end{center}
\end{figure}

For large values of beta there is still an early stage, lasting a few eddy turnover times, of turbulent stirring of the unstable initial condition. But this initial stirring is limited to scales smaller than the Rhines scale $L_{R}=2\pi E_{0}^{1/4}/\beta^{1/2}$, which eventually leads to a zonal PV field perturbed by Rossby waves around this scale, see Fig. \ref{fig:beta4}. The dynamics is then dominated by Rossby waves, just as in a single layer model \citet{Rhines77,Rhines75}. 


Because the perturbation is confined in the upper layer, irreversible turbulent stirring occurs in the upper layer only, but not at the bottom. We did check that the global distribution of coarse-grained PV levels did not change in the lower layer through time, while this distribution was changed in the upper layer. In this case, the fundamental assumption that the system explores evenly the available phase space through turbulent stirring is not valid. One might reasonably argue that the statistical mechanic theory is incomplete, for it cannot identify the phase space that can be explored. In any case, we  conclude that large values of beta prevent the convergence toward this equilibrium, leaving the question of the convergence itself to future work. 

\subsection{Turbulent stirring vs wave dynamics \label{sub:The-key-role-of-beta}}

\begin{figure}
\begin{center}
\includegraphics[width=0.9\textwidth]{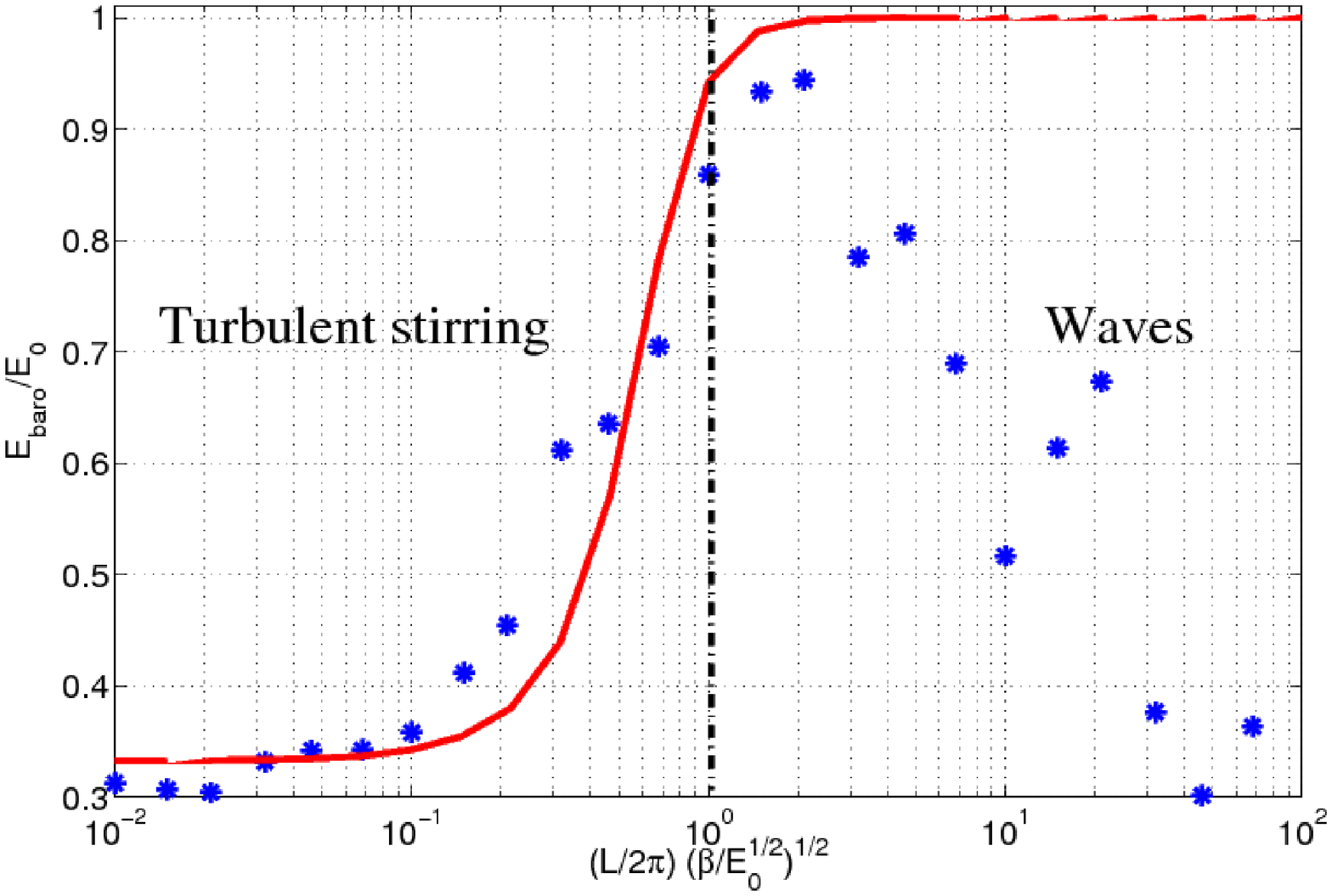}
\caption{{Variation of the Barotropic energy $E_{baro}$ normalized by the total energy $E_0$, varying the ratio of the domain scale with the Rhines scale $L/L_R=\beta^{1/2} E_{0}^{-1/4}L/2\pi$, at the end of each run, for the same initial condition $\psi_{0}$ (here $L=2\pi$, $E_{0}=1$, so only $\beta$ varies). Red continuous line: predictions from statistical mechanics. The vertical structure of the equilibrium state is computed with equation (\ref{eq:VertStruc}), using $K_\text{eff}=3.5$, and using the enstrophy profile $Z_{0}(\beta)$ of the initial condition $\psi_{0}$. The dashed-dotted line separates the wave regime from the turbulent stirring regime, in which we expect the statistical mechanics predictions to be valid.}  \label{fig:Ebaro_vs_beta}}
\end{center}
\end{figure}

We summarize the effect of planetary vorticity gradients on Fig.\
\ref{fig:Ebaro_vs_beta} by considering the ratio $E_\text{baro}/E_{0}$
of barotropic energy to total energy, after $250$ eddy turnover times,
for varying values of $\beta$. The plain red line represents statistical
mechanics predictions. In order to allow comparison with the numerical
experiments, for which the horizontal PV field did present structures
smaller than the gravest mode $K=1$ for low values of $\beta$ (see
subsection \ref{sub:SQG-like-dynamics-(small-beta)}), we computed
the vertical structure with equation (\ref{eq:VertStruc}), using
$K_\text{eff}=3.5$, and using the enstrophy profile $Z_{0}(\beta)$ of
the initial condition $\psi_{0}$. Strictly speaking, this effective
horizontal wavenumber $K_\text{eff}$ should vary with beta between
$K_\text{eff}=3.5$ for $\beta=0$ and $K_\text{eff}=1$ for  $\beta>1$, but taking into account this variations would not change much the shape of the red plain curve.

The critical value of $\beta$ between the turbulent stirring regime for which statistical mechanics predictions are useful and the wave regime can  be estimated by considering the case in which the Rhines scale $L_{Rh}\sim2\pi\left(E_{0}^{1/2}/\beta\right)^{1/2}$ is of the order of the domain scale $L=2\pi$. Here the total energy is $E_{0}=1$, which renders the critical value $\beta\sim1$ in our simulations.

\begin{figure}
\begin{center}
\includegraphics[width=\textwidth]{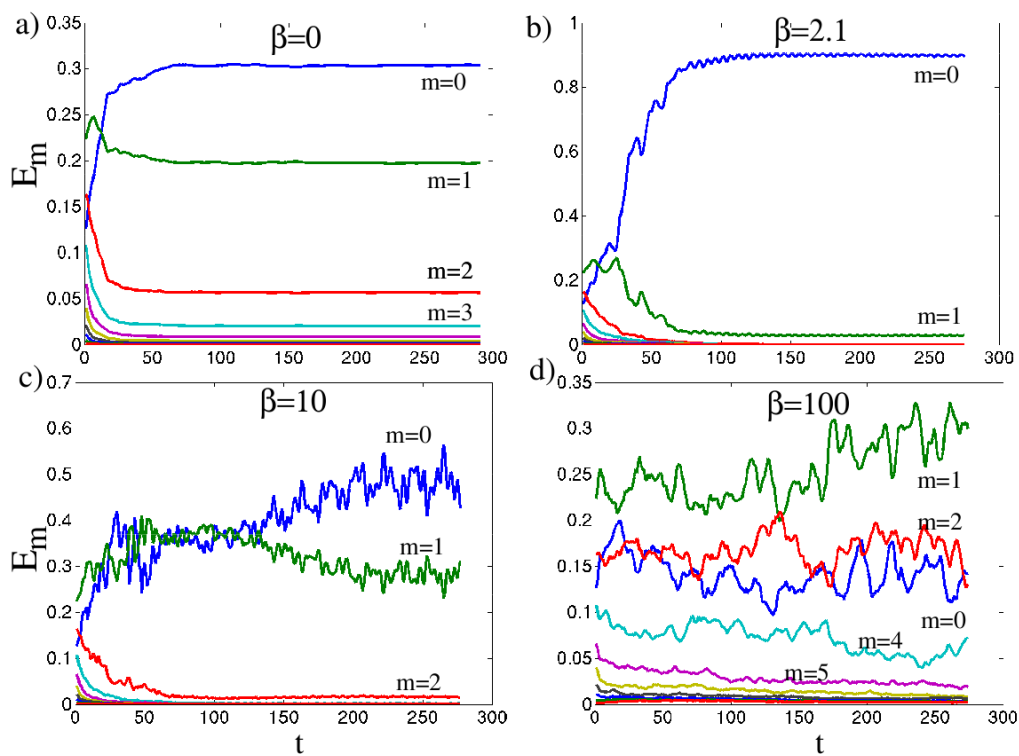}

\caption{Temporal evolution of the contribution of the energy of each baroclinic
mode to the total energy, for different values of $\beta$. The modes
are indexed by $m$, with $m=0$ the barotropic mode and $m=1$ the
first baroclinic mode. \label{fig:Emodes}}
\end{center}
\end{figure}

Because the ratio $E_\text{baro}/E_{0}$ gives incomplete information on
the flow structure, and because dynamical information is useful to
understand where statistical mechanics fails to predict
the vertical structure, we represent in Fig. \ref{fig:Emodes} the
temporal evolution of the energies $E_{m}$ of the baroclinic modes%
\footnote{Note that with this notation, $E_{m=0}=E_\text{baro}$ is the energy of
the barotropic mode, which is in general different from the total
energy of the initial condition $E_{0}$.%
} defined in (\ref{eq:VerticalModes}).

For small values of beta (see the first panel of Fig. \ref{fig:Emodes}),
the dynamics reaches the equilibrium after stirring during
few eddy turnover times, and the contribution of each mode $E_{m}$
simply reflects the SQG-like structure of the final state. 
For values of beta of order one (see the second panel of Fig. \ref{fig:Emodes}),
the same stirring mechanism leads to a barotropic flow after
a few eddy turnover times.
For values of $\beta$ larger than one, there is still an initial
 stirring regime, but the duration of this regime tends to
diminish with larger values of beta. We observe a slow transfer of
energy from first baroclinic to the barotropic mode in the case $\beta=10$
(see the third panel of Fig. \ref{fig:Emodes}), but it is unclear
if it will eventually lead to a condensation of the energy in the
barotropic mode. In the case $\beta=100$ (see the fourth panel of
Fig. \ref{fig:Emodes}), there is in average almost no energy transfers
between the vertical modes, after the initial stirring regime. Again,
it is unclear if interactions between waves would eventually condense
their energy into the barotropic mode in a longer numerical run. The
strong temporal fluctuations of the contribution of each vertical
mode visible on the wave regime of figure Fig. \ref{fig:Emodes} may
explain why they are large variations of $E_\text{baro}/E_{tot}$ (in addition
to the mean tendency of a decay with $\beta$) in the wave regime
 figure \ref{fig:Ebaro_vs_beta}. 

{To conclude, in the turbulent stirring regime (i.e. $L_R \sim L$ or $L_R>L$) MRS statistical mechanics predictions are qualitatively correct, and there are no ``transients'' (no temporal fluctuations) once the flow is self-organized. By contrast, in the wave regime (i.e. $L_R \ll L$) , there are strong temporal fluctuations in the mode amplitudes, and we believe that statistical mechanics (either MRS of energy-enstrophy equilibria for the truncated dynamics) can not account for these observations.}

\section{Discussions and conclusions \label{sec:Conclusion}}

We have used  statistical mechanics to predict the
flow structure resulting from the self-organization of freely evolving
inviscid stratified quasi-geostrophic flows. The only assumption of
the theory is that the flow explores evenly the available phase space.
In order to compute explicitly statistical equilibria, and to discuss
basic physical properties of these equilibria, we have focused on
MRS equilibria characterized by linear relation between potential
vorticity and streamfunction. For such MRS equilibrium states, we need the total energy and the fine-grained enstrophy in each layer, although all the invariants are implicitly taken into account.

We explained that these states are expected in a low energy limit
or in a strong mixing limit. By applying a method proposed by \citep{Bouchet:2008_Physica_D}, we have shown that MRS equilibria characterized by a linear $q-\psi$ relation are solutions of a minimum coarse-grained enstrophy variational
problem, which generalizes the Bretherton-Haidvogel minimum enstrophy
principle to the stratified case.

The central result of the paper is the elucidation of the physical
consequences of the conservation of the fine-grained enstrophy $Z_{0}(z)$
on the structure of the equilibrium states. We showed first that the
affine coefficients between the coarse-grained PV field and the streamfunction
are proportional to the fine-grained enstrophy: $\overline{q}=\beta_{t}Z_{0}(z)\psi$.
This relation allowed to compute explicitly the statistical equilibria,
and to discuss their structure depending on the profile $Z_{0}(z)$.
When this profile $Z_{0}$ is depth independent, the equilibrium state
is fully barotropic, whatever the stratification profile $N$. When
this profile $Z_{0}$ is surface intensified, the equilibrium state
is an SQG-like mode, characterized by an $e$-folding depth $\sim f_{0}/NK$,
where $1/K$ is bounded by the domain scale. Since the $e$-folding
depth increases with $K^{-1}$, larger horizontal flow structures
imply ``more barotropic'' flows. Since statistical equilibria
are associated with the gravest Laplacian horizontal modes ($K=1$),
the equilibrium state is the {}``most barotropic one'' given an
enstrophy profile $Z_{0}$. 

We conclude that the dynamics leads to the \emph{gravest
horizontal mode on the horizontal (inverse cascade), associated with
the gravest vertical structure consistent with the conservation of
the fine grained enstrophy. The flow becomes fully barotropic when
the fine-grained enstrophy profile is depth independent. }

These results can be used to understand the role of beta in barotropization in the following way: consider an initial SQG-like flow, with $\beta =0$, prescribed by its streamfunction; its enstrophy profile is surface intensified, and so will be the equilibrium sate. But when one switches on the beta effect, with the same initial streamfunction, the contribution of the depth independent part of the fine-grained enstrophy increases, which means a tendency toward a more barotropic equilibria, according to the previous conclusion. This result reflects the fact that in physical space, the initial SQG-like mode stirs the interior PV field (initially a beta plane), which in turn induces an interior flow, which stirs even more the interior PV field, and so on.

When the beta effect becomes large enough, the PV
distribution at each depth $z$ becomes prescribed by the initial
beta plane, so the most probable state should be also characterized
by a depth independent PV field, i.e. by a barotropic flow. 

To reach the equilibrium state, the PV field must be stirred
enough to explore the phase space. Yet large values of beta stabilize the initial condition, the system is trapped in a stable state different from the equilibrium state, and the flow dynamics becomes dominated by the interaction between Rossby waves rather than by turbulent stirring. The reason for this trapping is that the dynamics can not provide by itself sufficient perturbations to escape from the stable state and to explore other regions of the phase space, which would attract the system toward the equilibrium state. We observed in numerical simulations that this wave regime did not lead to a barotropization of the flow after a few hundred turnover times. We do not know if still longer numerical runs would lead to more barotropization. We estimated the critical value of beta between the turbulent stirring regime and the Rossby wave regime as the value such that the Rhines scale is of the order of the domain scale, which gives $\beta=4\pi^{2}E_{0}^{1/2}/L^{2}$.

{In this paper, we focused on the  effect of $\beta$ on barotropization process. More generally,  our results show that vertical transfers of energy and momentum are favored by the presence of any lateral potential vorticity gradients, since these gradients provide a  source of available perturbation enstrophy. These lateral PV gradients may for instance be due to large scale mean flows set by external process (large scale wind pattern for the oceans, and large scale temperature gradients for the atmosphere). A similar situation occurs in the framework of interactions between  waves (or eddies) and mean flows: horizontal momentum can be transfered vertically by inviscid form drag if there are horizontal fluxes of potential vorticity, which require themselves the presence of large scale potential vorticity gradients, see e.g. \cite{VallisBook}.}

{Another source of fine-grained potential vorticity would be provided by the addition of bottom topography. This should in this case play against barotropization, since the topography induces potential fine-grained enstrophy in the lower layer only. In fact, an initially surface intensified flow may evolve towards a bottom trapped current above the topographic anomaly, which can be explained by the statistical mechanics arguments presented in this paper, see \cite{Venaille12}.}\\

We now discuss how these results may apply to the mesoscale ocean, and in what respect they may provide interesting eddy parametrization. First, we have considered the relaxation of an initial condition by an inviscid flow. Real oceanic flows are forced dissipated. Statistical mechanics predictions should apply if a typical time scale for self-organization is smaller than typical time scale of forcing and dissipation. Issues of forcing and dissipation lead to another difficulty: what is the domain in which the flow self-organizes ? Strictly speaking, it should be the oceanic basin. But forcing and dissipation become dominant at basin scale (with, for instance, the Sverdrup balance setting the gyre structures). In practice, on a phenomenological basis, one could still apply the statistical mechanics predictions by considering an artificial domain scale with prescribed scale of arrest for the inverse cascade, governed say by dissipation processes, but which could not be predicted by statistical mechanics itself.

We  have also neglected bottom friction, because equilibrium statistical mechanics apply for non-dissipative systems. It is generally believed that bottom friction plays a key role in the vertical structure of oceanic flows \citet{ArbicFriction,ThompsonYoung2006}. On a phenomenological basis, quasi-statistical equilibria could be computed in the high bottom friction limit by adding a constraint of zero velocity at the bottom, for instance by considering $\psi(-H)=0$ in equation (\ref{eq:VertStruc}).

Third, equilibrium statistical mechanics predict the final state organization
of the freely evolving inviscid dynamics, but neither predicts
the route toward this state nor the corresponding typical time scales
for the convergence toward this state. Various parametrizations that relax
the flow toward the equilibrium state, following a path of maximum
entropy production have been proposed, see \citet{KazantsevSommeria98,RobertSommeria92}. These parametrization satisfy basic physical constraints satisfied by the dynamics (PV distribution conservation and energy conservation), but the actual dynamics may in some cases follow a different path than one of maximum entropy production. 

{For instance,  \citet{SmithVallis01,FuFlierl80} reported the existence of two time scales for the flow evolution in a freely evolving quasi-geostrophic turbulent flow with surface intensified stratification. On a fast time scale, the dynamics lead to a inverse cascade of vertical modes toward the first baroclinic modes, which is surface intensified, and observed the tendency toward barotropization at much larger time scale. The existence of these two time scales is a dynamical effect that can not be accounted by statistical mechanics.} 

The results presented in this paper have highlighted the important
role of the fine-grained enstrophy profile, for this imposes strong constraints
for the eddy structures. On a practical point of view, from the perspective
of eddy parametrizations, our result suggest that rather than assuming
a vertical structure of eddies given by SQG modes, or given by combination
of barotropic and first baroclinic modes, one could compute the eddy
structure with equation (\ref{eq:VertStruc}), assuming that $K$
is the scale of arrest, and $Z_{0}$ an enstrophy profile that could
be deduced from the resolved flow. For instance, $Z_{0}$ could be
set by the structure of the most unstable mode of local baroclinic
instabilities.

Observations indicate the presence of long-lived surface intensified
eddies in the ocean \citep{chelton,eddy1,eddy2}, and we here speculate as to how these eddies may be interpreted according to a statistical 
mechanical theory. First, the eddies may be interpreted as local
statistical equilibrium states of the continuously stratified dynamics
associated with a surface intensified fine-grained enstrophy profile,
much as we considered in the present paper. Second, they could be
interpreted as statistical equilibrium states of a 1.5 layer
quasi-geostrophic model assuming that only the upper layer of the
ocean is active, see e.g. \cite{VenailleBouchetJPO10}. Third, they
could be far from equilibrium states driven by physical mechanisms
preventing convergence towards statistical equilibrium (dynamical
effects due to non-uniform stratification, bottom friction, large
scale forcing). Understanding which hypothesis is relevant to explain
the formation of long-lived surface intensified ocean vortices will
require further research. 

\paragraph{Acknowledgments}
  This work was supported by DoE grant DE-SC0005189 and NOAA grant NA08OAR4320752. The authors sincerely thank Peter Rhines and two other reviewers for their very helpful comments provided during the review process. The authors also warmly thank F. Bouchet and J. Sommeria for fruitful discussions, and K.S. Smith for sharing his QG code. 

\appendix

\section{Minimum enstrophy states \label{sec:app1}}

The aim of this appendix is to show that the class of MRS statistical
associated with a linear $\overline{q}-\psi$ relation can be computed
by solving a minimum enstrophy variational problem satisfied by the
coarse-grained state $\overline{q}$. This is a direct consequence of a more general result of \citet{Bouchet:2008_Physica_D}, see also \citet{MajdaWangBook, NasoChavanisDubrulle2010}. {Here bottom topography is omitted to simplify the presentation, but taking into account this additional parameter is straightforward. In addition, the results obtained in this appendix are valid whatever the domain geometry. The only complication added by the presence of boundaries is that the circulation at these boundaries becomes a parameter that must be discussed.}

\subsection{Variational problem of the MRS statistical theory}

We consider the variational problem (\ref{eq:RSM}) given by MRS theory.
To compute its critical points, we introduce the Lagrange multipliers
$\beta_{t}$,$\alpha(\sigma,z),\zeta(x,y)$ associated respectively
with the energy $\mathcal{E}$ (\ref{eq:energy_functional_rho}),
with the global PV distribution $\mathcal{P}$ (\ref{eq:PV_constraint}),
and with the local normalization constraint $\mathcal{N}$ given by
(\ref{eq:normalization}). Critical points are solutions of\[
\delta\mathcal{S}-\beta_{t}\delta\mathcal{E}-\left(\int_{-H}^{0}\mathrm{d}z\ \int_{\Sigma}\mathrm{d}\sigma\ \alpha\delta\mathcal{P}\right)-\left(\int_{D}\mathrm{d}x\mathrm{d}y\ \zeta\delta\mathcal{N}\right)=0,\]
 where first variations are taken with respect to the field $\rho$.
This equation and the normalization constraint lead to 
\begin{equation}
\rho=\frac{1}{z_{\alpha}(\beta_{t}\psi)}\exp\left(\beta_{t}\psi\sigma-\alpha\left(\sigma,z\right)\right) \ , \text{with }  z_{\alpha}(u)=\int_{\Sigma}\mathrm{d}\sigma\ \exp\left(u\sigma-\alpha\left(\sigma,z\right)\right). \label{eq:rhoChar}
\end{equation}
 The coarse-grained potential vorticity field associated with
this state is

\begin{equation}
\overline{q}=\int_{\Sigma}\mathrm{d}\sigma\ \sigma\rho=f_{\alpha}\left(\beta_{t}\psi\right), \text{with } f_{\alpha}(u)=\frac{\mathrm{d}}{\mathrm{d}u}\ln z_{\alpha}; \  \label{eq:qmoy}\end{equation}
 and the local fluctuations of potential vorticity are

\begin{equation}
\overline{q^{2}}-\overline{q}^{2}=\int_{\Sigma}\mathrm{d}\sigma\ \sigma^{2}\rho=f_{\alpha}^{\prime}\left(\beta_{t}\psi\right),\label{eq:qfluc}\end{equation}
 where $f_{\alpha}^{\prime}$ is the derivative of $f_{\alpha}$.

\subsection{Case of a linear $q-\psi$ relation}

Here we focus on a particular class of solutions, assuming 

\[
f_{\alpha}(u)=\Omega(z)u,\]
 where the coefficient $\Omega(z)$ will be expressed in term of the
constraints of the problem. Equations (\ref{eq:qmoy}) and (\ref{eq:qfluc})
give the $q-\psi$ relation and the fluctuations

\[
\overline{q}=\beta_{t}\Omega(z)\psi,\quad\Omega=\overline{q^{2}}-\overline{q}^{2}.\]

Integrating the fluctuations on the horizontal leads to \begin{equation}
\Omega=\int_{\mathcal{D}}\mathrm{d}x\mathrm{d}y\ \left(\overline{q^{2}}-\overline{q}^{2}\right)=2\left(\mathcal{Z}-\mathcal{Z}_{cg}\right),  \text{with } \label{eq:AFluc}\end{equation}
\[ \mathcal{Z}[\rho]=\frac{1}{2}\int_{\mathcal{D}}\mathrm{d}x\mathrm{d}y\ \left(\int_{\Sigma}\mathrm{d}\sigma\ \sigma^{2}\rho\right),\quad\mathcal{Z}_{cg}[\rho]=\frac{1}{2}\int_{\mathcal{D}}\mathrm{d}x\mathrm{d}y\ \left(\int_{\Sigma}\mathrm{d}\sigma\ \sigma\rho\right)^{2}.\]
 Using the relations (\ref{eq:rhoChar}-\ref{eq:qmoy}),
we find that 

\begin{equation}
z_{\alpha}(u)=\left(2\pi\Omega\right)^{1/2}\exp\left(\frac{1}{2}\Omega u^{2}\right),\quad\rho=\frac{1}{\left(2\pi\Omega\right)^{1/2}}\exp\left(-\frac{\left(\sigma-\overline{q}\right)^{2}}{2\Omega}\right),\quad\alpha(\sigma,z)=\frac{\sigma^{2}}{2\Omega}.\label{eq:alphaFluc}
\end{equation}

\subsection{Link with a minimum enstrophy problem}

Solution of dual, less constrained variational problem is always a
solution of the more constrained problem for some values of the constraints
\citep{Ellis00}. We relax the constraints on the PV distribution
by introducing the Legendre Transform $\mathcal{G}_{\alpha}$ of the entropy
associated with this constraint, taking into account the conservation
of the total energy $E_{0}$ and of the fine-grained enstrophy $Z_{0}(z)$
in each layer:

\begin{equation}
G\left(E_{0},Z_{0}(z),\alpha\right)=\min_{\rho,\mathcal{N}[\rho]=1}\left\{ \mathcal{G}_{\alpha}\left[\rho\right]\ |\ \mathcal{E}[\rho]=E_{0},\ \mathcal{Z}[\rho]=Z_{0}(z)\right\} \,,\label{eq:RSMrelax_app}
\end{equation}

\begin{equation}
\mathcal{G}_{\alpha}=\int_{\mathcal{D}}\mathrm{d}x\mathrm{d}y\ \int_{-H}^{0}\mathrm{d}z\ \int\mathrm{d}\sigma\ \rho\ln\rho+\int_{-H}^{0}\mathrm{d}z\ \int_{\Sigma}\mathrm{d}\sigma\ \alpha\int_{\mathcal{D}}\mathrm{d}x\mathrm{d}y\ \rho,\label{eq:Grho}\end{equation}
where $\mathcal{E}$ is given by (\ref{eq:energy_functional_rho}),
where the pdf field is locally normalized using (\ref{eq:normalization}).
Equations (\ref{eq:alphaFluc}) and (\ref{eq:AFluc}) give \[
\alpha(\sigma,z)=\frac{\sigma^{2}}{4(Z_{0}-\mathcal{Z}_{cg})}\quad\text{and}\quad\rho=\frac{\exp\left(-\frac{\left(\sigma-\overline{q}\right)^{2}}{4(Z_{0}-\mathcal{Z}_{cg})}\right)}{\left(4\pi(Z_{0}-\mathcal{Z}_{cg})\right)^{1/2}}\]
 Injecting these expressions into (\ref{eq:Grho}) yields 
\[ \mathcal{G}_{\alpha}[\rho]=\frac{1}{2}\int_{-H}^{0} \mathrm{d}z\ w\left(\frac{\mathcal{Z}_{cg}}{Z_{0}}\right)+cst,\quad\text{with}\quad w(x)=-\ln(1-x)+\frac{1}{1-x}\]

In the limit $x \ll 1$, we have $w(x) \approx x$, so finding the minimizer $\rho_{m}$ of $\mathcal{G}_{\alpha}$ with the (global) constraint on the energy and the enstrophy is equivalent of finding the minimizer of the functional $\int_{-H}^0 \mathrm{d}z \ \mathcal{Z}_{cg}[\overline{q}]/Z_{0}$ with the same constraints. The hypothesis that $\mathcal{Z}_{cg} \ll Z_0$  is consistent with the ``strong mixing limit'' ($\beta_t \psi \sigma \ll 1$) or the low energy limit for which MRS equilibria are characterized by a linear relation between potential vorticity and streamfunction.  We conclude that MRS equilibrium states characterized by a linear $\overline{q}-\psi$ relation are solutions of the variational problem (\ref{eq:MinimumEnstrophyVP}).

\bibliographystyle{jfm}
\bibliography{AV,barotropization}

\begin{thebibliography}{51}
\expandafter\ifx\csname natexlab\endcsname\relax\def\natexlab#1{#1}\fi

\bibitem[{Arbic} \& {Flierl}(2004)]{ArbicFriction}
{\sc {Arbic}, B.~K. \& {Flierl}, G.~R.} 2004 {Baroclinically Unstable
  Geostrophic Turbulence in the Limits of Strong and Weak Bottom Ekman
  Friction: Application to Midocean Eddies}. {\em Journal of Physical
  Oceanography\/} {\bf 34}, 2257.

\bibitem[{Bouchet}(2008)]{Bouchet:2008_Physica_D}
{\sc {Bouchet}, F.} 2008 Simpler variational problems for statistical
  equilibria of the 2d euler equation and other systems with long range
  interactions. {\em Physica D Nonlinear Phenomena\/} {\bf 237}, 1976--1981.

\bibitem[{Bouchet} \& {Corvellec}(2010)]{BouchetCorvellec10}
{\sc {Bouchet}, F. \& {Corvellec}, M.} 2010 {Invariant measures of the 2D Euler
  and Vlasov equations}. {\em Journal of Statistical Mechanics: Theory and
  Experiment\/} {\bf 8}, 21.

\bibitem[{Bouchet} \& {Simonnet}(2009)]{BouchetSimonnetPRL09}
{\sc {Bouchet}, F. \& {Simonnet}, E.} 2009 {Random Changes of Flow Topology in
  Two-Dimensional and Geophysical Turbulence}. {\em Physical Review Letters\/}
  {\bf 102}~(9), 094504.

\bibitem[{Bouchet} \& {Venaille}(2012)]{BouchetVenaillePhysRep11}
{\sc {Bouchet}, F. \& {Venaille}, A.} 2012 {Statistical mechanics of two
  dimensional and geophysical turbulent flows}. {\em Physics Reports\/} {\bf
  515}, 227--295.

\bibitem[Bretherton(1966)]{Bretherton66}
{\sc Bretherton, F.P.} 1966 Critical layer instability in baroclinic flows.
  {\em Quart. J. Roy. Meteor. Soc.\/} {\bf 92}, 325--334.

\bibitem[{Bretherton} \& {Haidvogel}(1976)]{BrethertonHaidvogel}
{\sc {Bretherton}, F.~P. \& {Haidvogel}, D.~B.} 1976 {Two-dimensional
  turbulence above topography}. {\em Journal of Fluid Mechanics\/} {\bf 78},
  129--154.

\bibitem[{Campa} {\em et~al.\/}(2009){Campa}, {Dauxois} \&
  {Ruffo}]{CampaDauxoisRuffo}
{\sc {Campa}, A., {Dauxois}, T. \& {Ruffo}, S.} 2009 {Statistical mechanics and
  dynamics of solvable models with long-range interactions}. {\em Physics
  Reports\/} {\bf 480}, 57--159.

\bibitem[{Carnevale} \&
  {Frederiksen}(1987)]{Carnevale_Frederiksen_NLstab_statmech_topog_1987JFM}
{\sc {Carnevale}, G.~F. \& {Frederiksen}, J.~S.} 1987 {Nonlinear stability and
  statistical mechanics of flow over topography}. {\em Journal of Fluid
  Mechanics\/} {\bf 175}, 157--181.

\bibitem[{Chaigneau} {\em et~al.\/}(2011){Chaigneau}, {Le Texier}, {Eldin},
  {Grados} \& {Pizarro}]{eddy2}
{\sc {Chaigneau}, A., {Le Texier}, M., {Eldin}, G., {Grados}, C. \& {Pizarro},
  O.} 2011 {Vertical structure of mesoscale eddies in the eastern South Pacific
  Ocean: A composite analysis from altimetry and Argo profiling floats}. {\em
  Journal of Geophysical Research (Oceans)\/} {\bf 116}, 11025.

\bibitem[{Charney}(1971)]{Charney71}
{\sc {Charney}, J.~G.} 1971 {Geostrophic Turbulence.} {\em Journal of
  Atmospheric Sciences\/} {\bf 28}, 1087--1094.

\bibitem[{Chavanis} \&
  {Sommeria}(1996)]{ChavanisSommeria:1996_JFM_Classification}
{\sc {Chavanis}, P.~H. \& {Sommeria}, J.} 1996 {Classification of
  self-organized vortices in two-dimensional turbulence: the case of a bounded
  domain}. {\em J. Fluid Mech.\/} {\bf 314}, 267--297.

\bibitem[{Chelton} {\em et~al.\/}(2007){Chelton}, {Schlax}, {Samelson} \& {de
  Szoeke}]{chelton}
{\sc {Chelton}, D.~B., {Schlax}, M.~G., {Samelson}, R.~M. \& {de Szoeke},
  R.~A.} 2007 {Global observations of large oceanic eddies}. {\em Geophysical
  Research Letters\/} {\bf 34}, 15606.

\bibitem[{Corvellec}(2012)]{CorvellecPhD}
{\sc {Corvellec}, M.} 2012 {\em {PhD thesis, chapter 3}\/}. ENS-LYON,
  Universite Claude Bernard.

\bibitem[{Corvellec} \& {Bouchet}(2011)]{CorvellecBouchet2010}
{\sc {Corvellec}, M. \& {Bouchet}, F.} 2011 {A complete theory of low-energy
  phase diagrams for two-dimensional turbulence equilibria}. {\em preprint\/} .

\bibitem[{Ellis} {\em et~al.\/}(2000){Ellis}, {Haven} \& {Turkington}]{Ellis00}
{\sc {Ellis}, R.~S., {Haven}, K. \& {Turkington}, B.} 2000 Large deviation
  principles and complete equivalence and nonequivalence results for pure and
  mixed ensembles. {\em Journal of Statistical Physics\/} {\bf 101}, 999--1064.

\bibitem[{Ferrari} \& {Wunsch}(2009)]{FerrariWunsch09}
{\sc {Ferrari}, R. \& {Wunsch}, C.} 2009 {Ocean Circulation Kinetic Energy:
  Reservoirs, Sources, and Sinks}. {\em Annual Review of Fluid Mechanics\/}
  {\bf 41}, 253--282.

\bibitem[{Frederiksen}(1991{\natexlab{{\em a\/}}})]{FrederiksenB}
{\sc {Frederiksen}, J.~S.} 1991{\natexlab{{\em a\/}}} {Nonlinear stability of
  baroclinic flows over topography}. {\em Geophysical and Astrophysical Fluid
  Dynamics\/} {\bf 57}, 85--97.

\bibitem[{Frederiksen}(1991{\natexlab{{\em b\/}}})]{fredericksen91GAFDa}
{\sc {Frederiksen}, J.~S.} 1991{\natexlab{{\em b\/}}} {Nonlinear studies on the
  effects of topography on baroclinic zonal flows}. {\em Geophysical and
  Astrophysical Fluid Dynamics\/} {\bf 59}, 57--82.

\bibitem[{Frederiksen} \& {Sawford}(1980)]{FrederiksenSawford}
{\sc {Frederiksen}, J.~S. \& {Sawford}, B.~L.} 1980 {Statistical Dynamics of
  Two-Dimensional Inviscid Flow on a Sphere.} {\em Journal of Atmospheric
  Sciences\/} {\bf 37}, 717--732.

\bibitem[{Fu} \& {Flierl}(1980)]{FuFlierl80}
{\sc {Fu}, L. \& {Flierl}, G.} 1980 {Nonlinear energy and enstrophy transfers
  in a realistically stratified ocean}. {\em Dynamics of Atmospheres and
  Oceans\/} {\bf 4}, 219--246.

\bibitem[{Herbert} {\em et~al.\/}(2012){Herbert}, {Dubrulle}, {Chavanis} \&
  {Paillard}]{Corentin}
{\sc {Herbert}, C., {Dubrulle}, B., {Chavanis}, P.-H. \& {Paillard}, D.} 2012
  Phase transitions and marginal ensemble equivalence for freely evolving flows
  on a rotating sphere. {\em Physical Review E\/} {\bf 85}.

\bibitem[{Johnson} \& {McTaggart}(2010)]{eddy1}
{\sc {Johnson}, G.~C. \& {McTaggart}, K.~E.} 2010 {Equatorial Pacific 13C Water
  Eddies in the Eastern Subtropical South Pacific Ocean}. {\em Journal of
  Physical Oceanography\/} {\bf 40}, 226.

\bibitem[{Kazantsev} {\em et~al.\/}(1998){Kazantsev}, {Sommeria} \&
  {Verron}]{KazantsevSommeria98}
{\sc {Kazantsev}, E., {Sommeria}, J. \& {Verron}, J.} 1998 {Subgrid-Scale Eddy
  Parameterization by Statistical Mechanics in a Barotropic Ocean Model}. {\em
  Journal of Physical Oceanography\/} {\bf 28}, 1017--1042.

\bibitem[{Kraichnan}(1967)]{Kraichnan_1967PhFl...10.1417K}
{\sc {Kraichnan}, R.~H.} 1967 {Inertial Ranges in Two-Dimensional Turbulence}.
  {\em Physics of Fluids\/} {\bf 10}, 1417--1423.

\bibitem[{Kraichnan} \&
  {Montgomery}(1980)]{Kraichnan_Motgommery_1980_Reports_Progress_Physics}
{\sc {Kraichnan}, R.~H. \& {Montgomery}, D.} 1980 {Two-dimensional turbulence}.
  {\em Reports on Progress in Physics\/} {\bf 43}, 547--619.

\bibitem[{Lapeyre}(2009)]{LapeyreJPO09}
{\sc {Lapeyre}, G.} 2009 {What Vertical Mode Does the Altimeter Reflect? On the
  Decomposition in Baroclinic Modes and on a Surface-Trapped Mode}. {\em
  Journal of Physical Oceanography\/} {\bf 39}, 2857.

\bibitem[Lapeyre \& Klein(2006)]{LapeyreKlein06}
{\sc Lapeyre, G. \& Klein, P.} 2006 Dynamics of the upper oceanic layers in
  terms of surface quasigeostrophy theory. {\em Journal of Physical
  Oceanography\/} {\bf 36}, 165--176.

\bibitem[{Majda} \& {Wang}(2006)]{MajdaWangBook}
{\sc {Majda}, A. \& {Wang}, X.} 2006 {\em {Nonlinear Dynamics and Statistical
  Theories for Basic Geophysical Flows}\/}.

\bibitem[{McWilliams} {\em et~al.\/}(1994){McWilliams}, {Weiss} \&
  {Yavneh}]{McWilliams94}
{\sc {McWilliams}, J.~C., {Weiss}, J.~B. \& {Yavneh}, I.} 1994 {Anisotropy and
  Coherent Vortex Structures in Planetary Turbulence}. {\em Science\/} {\bf
  264}, 410--413.

\bibitem[{Merryfield}(1998)]{Merryfield98JFM}
{\sc {Merryfield}, W.~J.} 1998 {Effects of stratification on quasi-geostrophic
  inviscid equilibria}. {\em Journal of Fluid Mechanics\/} {\bf 354}, 345--356.

\bibitem[{Michel} \& {Robert}(1994)]{Michel_Robert_LargeDeviations94C}
{\sc {Michel}, J. \& {Robert}, R.} 1994 {Large deviations for young measures
  and statistical mechanics of infinite dimensional dynamical systems with
  conservation law}. {\em Communications in Mathematical Physics\/} {\bf 159},
  195--215.

\bibitem[Miller(1990)]{Miller:1990_PRL_Meca_Stat}
{\sc Miller, Jonathan} 1990 Statistical mechanics of euler equations in two
  dimensions. {\em Phys. Rev. Lett.\/} {\bf 65}~(17), 2137--2140.

\bibitem[{Naso} {\em et~al.\/}(2010){Naso}, {Chavanis} \&
  {Dubrulle}]{NasoChavanisDubrulle2010}
{\sc {Naso}, A., {Chavanis}, P.~H. \& {Dubrulle}, B.} 2010 {Statistical
  mechanics of two-dimensional Euler flows and minimum enstrophy states}. {\em
  European Physical Journal B\/} {\bf 77}, 187--212.

\bibitem[{Pedlosky}(1982)]{PedloskyBook}
{\sc {Pedlosky}, J.} 1982 {\em {Geophysical fluid dynamics}\/}.

\bibitem[{Rhines}(1977)]{Rhines77}
{\sc {Rhines}, P.}, ed. 1977 {\em The dynamics of unsteady currents\/}, ,
  vol.~6. Wiley and Sons.

\bibitem[Rhines(1975)]{Rhines75}
{\sc Rhines, P.~B.} 1975 Waves and turbulence on a $\beta$-plane. {\em J.
  Fluid. Mech.\/} {\bf 69}, 417--443.

\bibitem[{Robert}(1990)]{Robert:1990_CRAS}
{\sc {Robert}, R.} 1990 {Etats d'equilibre statistique pour l'ecoulement
  bidimensionnel d'un fluide parfait}. {\em C. R. Acad. Sci.\/} {\bf 1},
  311:575--578.

\bibitem[{Robert} \& {Sommeria}(1991)]{SommeriaRobert:1991_JFM_meca_Stat}
{\sc {Robert}, R. \& {Sommeria}, J.} 1991 {Statistical equilibrium states for
  two-dimensional flows}. {\em J. Fluid Mech.\/} {\bf 229}, 291--310.

\bibitem[{Robert} \& {Sommeria}(1992)]{RobertSommeria92}
{\sc {Robert}, R. \& {Sommeria}, J.} 1992 {Relaxation towards a statistical
  equilibrium state in two-dimensional perfect fluid dynamics}. {\em Physical
  Review Letters\/} {\bf 69}, 2776--2779.

\bibitem[{Salmon}(1998)]{Salmon_1998_Book}
{\sc {Salmon}, R.} 1998 {\em {Lectures on Geophysical Fluid Dynamics}\/}.
  Oxford University Press.

\bibitem[{Salmon} {\em et~al.\/}(1976){Salmon}, {Holloway} \&
  {Hendershott}]{SalmonHollowayHendershott:1976_JFM_stat_mech_QG}
{\sc {Salmon}, R., {Holloway}, G. \& {Hendershott}, M.~C.} 1976 {The
  equilibrium statistical mechanics of simple quasi-geostrophic models}. {\em
  Journal of Fluid Mechanics\/} {\bf 75}, 691--703.

\bibitem[{Schecter}(2003)]{SchecterPRE03}
{\sc {Schecter}, D.~A.} 2003 {Maximum entropy theory and the rapid relaxation
  of three-dimensional quasi-geostrophic turbulence}. {\em Physical Review E\/}
  {\bf 68}~(6), 066309.

\bibitem[{Scott} \& {Wang}(2005)]{ScottWang05}
{\sc {Scott}, R.~B. \& {Wang}, F.} 2005 {Direct Evidence of an Oceanic Inverse
  Kinetic Energy Cascade from Satellite Altimetry}. {\em Journal of Physical
  Oceanography\/} {\bf 35}, 1650.

\bibitem[{Smith} \& {Vallis}(2001)]{SmithVallis01}
{\sc {Smith}, K.~S. \& {Vallis}, G.~K.} 2001 {The Scales and Equilibration of
  Midocean Eddies: Freely Evolving Flow}. {\em Journal of Physical
  Oceanography\/} {\bf 31}, 554--571.

\bibitem[{Thomas} {\em et~al.\/}(2008){Thomas}, {Tandon} \& A.]{LeithThomas}
{\sc {Thomas}, L.~N., {Tandon}, A. \& A., {Mahadevan}} 2008 Sub-mesoscale
  processes and dynamics. {\em Eddy Resolving Ocean Modeling, M. W. Hecht and
  H. Hasumi eds. Amer. Geophys. Union\/} pp. 17--38.

\bibitem[{Thompson} \& {Young}(2006)]{ThompsonYoung2006}
{\sc {Thompson}, A.~F. \& {Young}, W.~R.} 2006 {Scaling Baroclinic Eddy Fluxes:
  Vortices and Energy Balance}. {\em Journal of Physical Oceanography\/} {\bf
  36}, 720.

\bibitem[{Vallis}(2006)]{VallisBook}
{\sc {Vallis}, G.~K.} 2006 {\em {Atmospheric and Oceanic Fluid Dynamics}\/}.

\bibitem[{Venaille}(2012)]{Venaille12}
{\sc {Venaille}, A.} 2012 Bottom trapped currents as statistical equilibrium
  states above topographic anomalies. {\em Journal of Fluid Mechanics\/} {\bf
  699}.

\bibitem[{Venaille} \& {Bouchet}(2011{\natexlab{{\em
  a\/}}})]{VenailleBouchetJPO10}
{\sc {Venaille}, A. \& {Bouchet}, F.} 2011{\natexlab{{\em a\/}}} {Ocanic rings
  and jets as statistical equilibrium states}. {\em Journal of Physical
  Oceanography\/} {\bf 41}, 1860,1873.

\bibitem[{Venaille} \& {Bouchet}(2011{\natexlab{{\em
  b\/}}})]{VenailleBouchetJSP11}
{\sc {Venaille}, A. \& {Bouchet}, F.} 2011{\natexlab{{\em b\/}}} {Solvable
  Phase Diagrams and Ensemble Inequivalence for Two-Dimensional and Geophysical
  Turbulent Flows}. {\em Journal of Statistical Physics\/} {\bf 143}, 346--380.

\end{thebibliography}

\end{document}